\documentclass[twocolumn,amsfonts,prd,nofootinbib,showpacs,10pt]{revtex4-1}
\usepackage{textcomp}
\usepackage{amsmath}
\usepackage{amsfonts}
\usepackage{calc}
\usepackage{mathrsfs}
\usepackage{amssymb}
\usepackage{amsmath}
\usepackage{array}
\usepackage{color}
\usepackage{bm}
\usepackage{graphicx}
\usepackage{hyperref}

\newcommand{\balpha}{{\bar\alpha}}
\newcommand{\bbeta}{{\bar\beta}}
\newcommand{\bgamma}{{\bar\gamma}}
\newcommand{\bdelta}{{\bar\delta}}
\newcommand{\bmu}{{\bar\mu}}
\newcommand{\bnu}{{\bar\nu}}
\newcommand{\bsigma}{{\bar\sigma}}

\newcommand{\bu}{{\bar u}}
\renewcommand{\r}{{\sf r}}
\newcommand{\s}{{\sf s}}
\renewcommand{\P}{\mathcal{P}}
\newcommand{\res}{\mathcal{R}}
\renewcommand{\S}{{\rm S}}
\newcommand{\R}{{\rm R}}

\newcommand{\e}{\epsilon}
\newcommand{\nhat}{\hat{n}}
\newcommand{\E}{\mathcal{E}}
\newcommand{\B}{\mathcal{B}}

\newcounter{subeq}

\begin{document}
\title{A practical, covariant puncture for second-order self-force calculations} 
\author{Adam Pound and Jeremy Miller} 
\affiliation{Mathematical Sciences, University of Southampton, Southampton, United Kingdom, SO17 1BJ}
\date{\today}

\begin{abstract}
Accurately modeling an extreme-mass-ratio inspiral requires knowledge of the second-order gravitational self-force on the inspiraling small object. Recently, numerical puncture schemes have been formulated to calculate this force, and their essential analytical ingredients have been derived from first principles. However, the \emph{puncture}, a local representation of the small object's self-field, in each of these schemes has been presented only in a local coordinate system centered on the small object, while a numerical implementation will require the puncture in coordinates covering the entire numerical domain. In this paper we provide an explicit covariant self-field as a local expansion in terms of Synge's world function. The self-field is written in the Lorenz gauge, in an arbitrary vacuum background, and in forms suitable for both self-consistent and Gralla-Wald-type representations of the object's trajectory. We illustrate the local expansion's utility by sketching the procedure of constructing from it a numerically practical puncture in any chosen coordinate system.
\end{abstract}
\maketitle 

\section{Introduction}
Observation of extreme-mass-ratio inspirals (EMRIs) is a central plank in plans for a space-based gravitational-wave detector~\cite{eLISA:13}. EMRIs, in which a compact object of mass $m$ orbits about and eventually falls into a massive black hole of mass $M\gg m$, will offer a unique probe of strong-field dynamics and a detailed map of the spacetime geometry near a black hole. However, an inspiral occurs on the very long dynamical timescale $M^2/m$, and to extract information about an inspiral from an observed waveform, one will require a model that accurately relates the waveform to the motion over that long time. For a physically relevant mass ratio $m/M=10^{-6}$, this translates to requiring an accurate model covering $\sim\!\!10^6$ wavecycles. 

Because of the drastically dissimilar lengthscales in these systems, numerical relativity cannot adequately model them even on short timescales. And because of the strong fields and large velocities in play, post-Newtonian theory is inapplicable. Instead, the most prominent method of tackling the problem has been to apply the gravitational self-force formalism~\cite{Barack:09,Poisson-Pound-Vega:11}, in which the small object is treated as the source of a perturbation $h_{\mu\nu}\sim m$ on the background spacetime $g_{\mu\nu}$ of the large black hole, and $h_{\mu\nu}$ exerts a force back on the small object, accelerating it away from test-particle, geodesic motion in $g_{\mu\nu}$. It has long been known~\cite{Rosenthal:06a} that within this formalism, accurately modeling an inspiral on the long timescale $\sim M^2/m$ requires knowledge of the smaller object's acceleration to second order in $m$, meaning garden-variety linear perturbation theory is insufficient. The veracity of this claim can be seen from a simple scaling argument: if the small object's acceleration contains an error of order $\delta a~\sim m^2/M^3$, then after a time $M^2/m$ the error in its position is $\delta z\sim t^2\delta a\sim M$ (setting $c=G=1$, as we do throughout this paper). Therefore, to ensure that the errors remain small (i.e., $\delta z\ll M$), we must allow no error in the acceleration at order $m^2$. In other words, we must account for the second-order self-force.\footnote{A subtler scaling argument~\cite{Hinderer-Flanagan:08} shows that only a specific piece of the second-order force is needed: the orbit-averaged dissipative piece, which causes the largest long-term changes in the orbit.}

In addition to its applications in the EMRI problem, the second-order self-force promises to be a useful tool in modeling other binary systems. At first order, numerical self-force data has been fruitfully used to fix high-order terms and otherwise-free parameters in post-Newtonian~\cite{Blanchet-etal:10b,Favata:11,LeTiec-etal:11,LeTiec-etal:12b} and effective-one-body~\cite{Damour:09,Barack-Damour-Sago:10, Barausse-etal:11,Akcay-etal:12} models, and the same strategy could be employed at second order. Perhaps more strikingly, at first order there is compelling evidence that the self-force formalism can be made accurate well outside the extreme-mass-ratio regime~\cite{LeTiec-etal:11,LeTiec-etal:13}, which suggests that at second order the self-force could be used to directly model intermediate-mass-ratio and potentially even comparable-mass binaries with reasonable accuracy.

After several exploratory studies of the second-order problem~\cite{Rosenthal:06a,Rosenthal:06b,Pound:10a,Detweiler:11}, these prospects have recently been brought substantially closer to realization, and the essential analytical ingredients necessary for concrete calculations of the second-order self-force are now available~\cite{Pound:12a,Gralla:12,Pound:12b,Pound:14a}. These ingredients are
\begin{itemize}
\item a local expression for the small object's \emph{self-field} $h^\S_{\mu\nu}$,
\item an equation of motion for the small object's center of mass in terms of a certain \emph{effective field} $h^\R_{\mu\nu}$.
\end{itemize} 
Both results were derived from the Einstein equations via rigorous methods of matched asymptotic expansions developed in Refs.~\cite{Gralla-Wald:08, Pound:10a}; for an overview, see the review~\cite{Poisson-Pound-Vega:11} or the forthcoming exegesis~\cite{Pound:14b}. Together, the above two ingredients make up all the requisite input for a numerical puncture scheme (also known as an effective-source scheme)~\cite{Barack-Golbourn:07,Vega-Detweiler:07,Wardell-etal:11}. In the context of matched asymptotic expansions, such a scheme originates from a split of the full perturbation $h_{\mu\nu}$ into two pieces:
\begin{equation}
h_{\mu\nu}=h^\S_{\mu\nu}+h^\R_{\mu\nu},
\end{equation}
where the self-field $h^\S_{\alpha\beta}$ encapsulates local information about the object's multipole structure, and the effective field $h^\R_{\mu\nu}$ is a vacuum perturbation that is determined by the global boundary conditions imposed on $h_{\mu\nu}$. 

$h^\S_{\mu\nu}$ and $h^\R_{\mu\nu}$ are defined locally in a neighbourhood \emph{outside} the object. A puncture scheme (and from this perspective, \emph{every} numerical scheme that has been implemented in calculations of the gravitational self-force) proceeds by analytically continuing these fields into the region where the object would lie in the full, physical spacetime. The analytically continued self-field $h^\S_{\mu\nu}$ diverges at a worldline $\gamma$ that represents the object's mean motion in the background spacetime, and the self-field is hence renamed the \emph{singular field}. Conversely, the analytically continued field $h^\R_{\mu\nu}$ is smooth at $\gamma$, earning it the sobriquet \emph{regular field}. In this paper we will use the choice of $h^\R_{\mu\nu}$ and $h^\S_{\mu\nu}$ defined by Pound~\cite{Pound:12a},\footnote{This definition is closely related to but slightly different from that of Ref.~\cite{Pound:12b}; see Sec.~\ref{singular-regular} and Appendix~\ref{trace-reverse}.} described again below in Sec.~\ref{Fermi-field}. With that choice, the effective metric $g_{\mu\nu}+h^\R_{\mu\nu}$ is a $C^\infty$ solution to the vacuum Einstein equation, and $\gamma$ is a geodesic in that vacuum metric (through order $m^2$, for any object with sufficient sphericity and slow spin). Alternative choices of $h^\R_{\mu\nu}$ and $h^\S_{\mu\nu}$, with different properties, have also been made at second order~\cite{Harte:12,Gralla:12}, and they could equally well be used within a puncture scheme.

Once the choice of singular and regular fields has been made, a puncture scheme begins with the construction of a \emph{puncture} $h^\P_{\mu\nu}$, defined by truncating a local expansion of the singular field, in powers of spatial distance $\lambda$ from $\gamma$, at a specified order. One then defines the residual field 
\begin{equation}
h^\res_{\mu\nu} \equiv h_{\mu\nu}-h^\P_{\mu\nu}
\end{equation}
and in a region covering the object, writes a field equation for $h^\res_{\mu\nu}$, rather than one for (the analytically continued) physical field $h_{\mu\nu}$. Since $h^\P_{\mu\nu}\approx h^\S_{\mu\nu}$, so too $h^\res_{\mu\nu}\approx h^\R_{\mu\nu}$. The better $h^\P_{\mu\nu}$ represents $h^\S_{\mu\nu}$, the better $h^\res_{\mu\nu}$ represents $h^R_{\mu\nu}$. For example, if $\lim_{x\to\gamma}[h^\P_{\mu\nu}(x)-h^\S_{\mu\nu}(x)]=0$, then $\lim_{x\to\gamma}h^\res_{\mu\nu}(x)=\lim_{x\to\gamma}h^\R_{\mu\nu}(x)$; that is, the residual field agrees with the regular field on the worldline. If $h^\P_{\mu\nu}$ is one order more accurate, meaning $h^\P_{\mu\nu}-h^\S_{\mu\nu}=o(\lambda)$, then $\lim_{x\to\gamma}\nabla_{\!\rho} h^\res_{\mu\nu}=\lim_{x\to\gamma}\nabla_{\!\rho} h^\R_{\mu\nu}$; since the self-force is constructed from first derivatives of $h^\R_{\mu\nu}$, this condition guarantees that the force can be calculated from $h^\res_{\mu\nu}$, as in Eq.~\eqref{motion_SC} below.\footnote{The reader should note that up to numerical error, this procedure yields the force {\em exactly}. No approximations are made by replacing the regular field with the residual field in the equation of motion.} 

This type of scheme removes the physical system in the interior of the object, with all its matter fields, curvature singularities  (in the case of a black hole), or even wormholes, and replaces it with an \emph{effective} system. Put more simplistically, the puncture replaces the object. The precise form that a puncture scheme takes, and the interpretation of the puncture's `position', will depend on the type of perturbative expansion one begins from: a self-consistent expansion~\cite{Pound:10a,Pound:12a,Pound:12b,Pound:13a,Pound:14b}; or what we will call a Gralla-Wald-type expansion, exemplified by Refs.~\cite{Gralla-Wald:08,Gralla:12}. The core difference between these two methods of expansion is their representation of the object's mean motion, but that difference influences the overarching treatment of the field equations. To set the stage for our calculations and establish a unified framework for the discussion, we will briefly describe the second-order puncture scheme that arises from each of the methods.

\subsection{Self-consistent puncture scheme}
In a self-consistent expansion of the field equations, one seeks an equation of motion for a self-accelerated worldline $\gamma$ that well represents the object's bulk motion, and one expands the metric perturbation in terms of functionals of that worldline:
\begin{equation}\label{h_SC_expansion}
h_{\mu\nu} = \e h^1_{\mu\nu}[\gamma]+\e^2 h^2_{\mu\nu}[\gamma]+O(\e^3),
\end{equation}
where $\e\equiv1$ is used to count powers of the object's mass $m$. Here each $h^n_{\mu\nu}$ is allowed a functional dependence on the accelerated (and therefore $\e$-dependent) worldline $\gamma$. After imposing the Lorenz gauge on the full perturbation,
\begin{equation}
\nabla^\nu\bar h_{\mu\nu} = 0,\label{gauge}
\end{equation}
where $\bar h_{\mu\nu} \equiv h_{\mu\nu}-\frac{1}{2}g_{\mu\nu}g^{\alpha\beta}h_{\alpha\beta}$, the vacuum Einstein equation outside the object, $R_{\mu\nu}[g+h]$, is split into a sequence of wave equations, the first two of which read
\begin{align}
E_{\mu\nu}[h^{1}] &= 0,\label{h1_eq}\\
E_{\mu\nu}[h^2] &= 2\delta^2R_{\mu\nu}[h^1,h^1],\label{h2_eq}
\end{align}
where
\begin{equation}
E_{\mu\nu}[h] \equiv \Box h_{\mu\nu} +2R_\mu{}^\alpha{}_\nu{}^\beta h_{\alpha\beta}
\end{equation}
is the usual tensorial wave operator and
\begin{align}
\delta^2R_{\alpha\beta}[h,h] &=
					-\tfrac{1}{2}\bar h^{\mu\nu}{}_{;\nu}\left(2h_{\mu(\alpha;\beta)}
					-h_{\alpha\beta;\mu}\right) 
					\nonumber\\
					&\quad +\tfrac{1}{4}h^{\mu\nu}{}_{;\alpha}h_{\mu\nu;\beta} 
					+\tfrac{1}{2}h^{\mu}{}_{\beta}{}^{;\nu}\left(h_{\mu\alpha;\nu} -h_{\nu\alpha;\mu}\right)\nonumber\\
					&\quad-\tfrac{1}{2}h^{\mu\nu}\left(2h_{\mu(\alpha;\beta)\nu}-h_{\alpha\beta;\mu\nu}-h_{\mu\nu;\alpha\beta}\right)\!\!\label{second-order_Ricci}
\end{align}
is the second-order Ricci tensor (the first term of which, involving $\bar h^{\mu\nu}{}_{;\nu}$, vanishes with our choice of gauge). Both a semicolon and $\nabla$ denote the covariant derivative compatible with the background metric $g_{\mu\nu}$.

After solving Eqs.~\eqref{h1_eq} and \eqref{h2_eq} in a region around the small object, each $h^n_{\mu\nu}$ can be decomposed into singular and regular pieces, or into a puncture and residual field:
\begin{align}
h^1_{\mu\nu} &= h^{\S1}_{\mu\nu}[\gamma] + h^{\R1}_{\mu\nu}[\gamma]= h^{\P1}_{\mu\nu}[\gamma] + h^{\res1}_{\mu\nu}[\gamma],\\
h^2_{\mu\nu} &= h^{\S2}_{\mu\nu}[\gamma] + h^{\R2}_{\mu\nu}[\gamma]= h^{\P2}_{\mu\nu}[\gamma] + h^{\res2}_{\mu\nu}[\gamma].
\end{align}
For a sufficiently slowly spinning object, the first- and second-order singular fields (and punctures) near $\gamma$ have the schematic forms
\begin{equation}\label{hS1_SC_schematic}
h^{\S1}_{\mu\nu}\sim \frac{m}{|x^i-z^i|} + O(|x^i-z^i|^0)
\end{equation}
and
\begin{equation}\label{hS2_SC_schematic}
h^{\S2}_{\mu\nu}\sim \frac{m^2}{|x^i-z^i|^2} + \frac{\delta m_{\mu\nu}+mh^{\R1}_{\mu\nu}}{|x^i-z^i|} + O(\ln|x^i-z^i|),
\end{equation}
where $z^i$ are spatial coordinates on $\gamma$ and $|x^i-z^i|$ represents distance from $\gamma$. The explicit expressions for the first few terms in these expansion, derived in Refs.~\cite{Pound:10a,Pound:12a}, are given in Eqs.~\eqref{hS1_SC_Fermi} and \eqref{hS2_SC_Fermi}--\eqref{hdm_SC_Fermi} in a local coordinate system $(t,x^i)$ centered on $\gamma$ (such that $z^i\equiv0$ in the schematic expressions above). At first order, the puncture is given roughly by a Coulomb potential sourced by the mass $m$. At second order, there are naturally quadratic combinations of this potential, signified by the $m^2$ term, but there are also quadratic combinations of the mass and the first-order regular field, as well as a gravitationally induced correction to the body's monopole moment, denoted by $\delta m_{\mu\nu}$ and given explicitly in Eq.~\eqref{dm_SC_Fermi}.

There are several schemes that can be developed from the starting point of the puncture. Here we describe a worldtube scheme in the tradition of Refs.~\cite{Barack-Golbourn:07,Dolan-Barack:11}. In this type of scheme one uses the field variables $h^{\res n}_{\mu\nu}$ inside a worldtube $\Gamma$ surrounding $\gamma$, the field variables $h^n_{\mu\nu}$ outside that worldtube, and the change of variables $h^n_{\mu\nu}=h^{\res n}_{\mu\nu}+h^{\P n}_{\mu\nu}$ when moving between the two regions.\footnote{One does not solve the problem in each domain separately, since the separate problems would be ill-posed. Instead, when calculating $h^n_{\mu\nu}$ at a point just outside $\Gamma$ that depends on points on past time slices inside $\Gamma$, one makes use of the values of $h^{\res n}_{\mu\nu}$ already calculated at those earlier points, and vice versa; see Sec.~VB of Ref.~\cite{Barack-Golbourn:07}.} The second-order puncture scheme is then summarized by the coupled system of equations
\begin{subequations}\label{h1_SC}%
\begin{align}
E_{\mu\nu}[h^{\res1}] &= -16\pi \bar T^1_{\mu\nu}[\gamma]-E_{\mu\nu}[h^{\P1}] & \text{inside }\Gamma,\\
E_{\mu\nu}[h^{1}] &= 0 & \text{outside }\Gamma,
\end{align}
\vspace{-1.5\baselineskip}
\end{subequations}
\begin{subequations}\label{h2_SC}%
\begin{align}
E_{\mu\nu}[h^{\res2}] &= 2\delta^2R_{\mu\nu}[h^1,h^1]-16\pi \bar T^2_{\mu\nu}[\gamma]\hspace{-15pt} &  \nonumber\\
											&\quad - E_{\mu\nu}[h^{\P2}] & \text{inside }\Gamma,\\
E_{\mu\nu}[h^2] &= 2\delta^2R_{\mu\nu}[h^1,h^1] & \text{outside }\Gamma,
\end{align}
\end{subequations}
\vspace{-1.5\baselineskip}
\begin{align}
\frac{D^2 z^\mu}{d\tau^2} &= -\frac{1}{2}P^{\mu\nu}\left(g_\nu{}^\gamma-h^\res_\nu{}^\gamma\right)
		\left(2h^\res_{\gamma\alpha;\beta}-h^\res_{\alpha\beta;\gamma}\right)u^\alpha u^\beta,\label{motion_SC}
\end{align}
where the puncture diverges on the worldline $z^\mu$ determined by Eq.~\eqref{motion_SC}. That divergence is quite strong, with the terms $2\delta^2R_{\mu\nu}[h^1,h^1]$ and $E_{\mu\nu}[h^{\P2}]$ in Eq.~\eqref{h2_SC} each blowing up as $1/\lambda^4$ near the worldline, but by construction, the divergences necessarily cancel each other.

Here the quantities
\begin{align}
\bar T^1_{\mu\nu}[\gamma] &= \int_\gamma m(\tfrac{1}{2}g_{\mu\nu}+u_\mu u_\nu)\delta^4(x,z)d\tau,\label{T1_SC}\\
\bar T^2_{\mu\nu}[\gamma] &= \int_\gamma \tfrac{1}{4}\delta m_{\mu\nu}\delta^4(x,z)d\tau,\label{T2_SC}
\end{align}
with $\delta^4(x,z)\equiv\delta^4(x^\alpha-z^\alpha)/\sqrt{-g}$, are effective (trace-reversed) point-particle stress-energy tensors sourcing the Coulomb-like fields $m/|x^i-z^i|$ and $\delta m_{\mu\nu}/|x^i-z^i|$ in Eqs.~\eqref{hS1_SC_schematic} and \eqref{hS2_SC_schematic}. Their origin is described in Sec.~\ref{Fermi-field} below. In the equation of motion~\eqref{motion_SC},
\begin{equation}
h^\res_{\mu\nu} = \e h^{\res1}_{\mu\nu}[\gamma]+\e^2 h^{\res2}_{\mu\nu}[\gamma]
\end{equation}
is the total residual field through second order, $\tau$ is proper time (measured in $g_{\mu\nu}$) on $\gamma$, $u^\mu\equiv \frac{dz^\mu}{d\tau}$ is the four-velocity on $\gamma$, $\frac{D}{d\tau}\equiv u^\mu\nabla_{\!\mu}$ is a covariant derivative along $u^\mu$, and
\begin{equation}
P^{\mu\nu}\equiv g^{\mu\nu}+u^\mu u^\nu
\end{equation}
projects orthogonally to $u^\mu$.

In this scheme, Eqs.~\eqref{h1_SC}--\eqref{motion_SC} must be solved together, as a coupled system for the variables $z^\mu$, $h^{\res1}_{\mu\nu}\slash h^1_{\mu\nu}$ (inside/outside $\Gamma$), and $h^{\res2}_{\mu\nu}\slash h^2_{\mu\nu}$. Unlike in many approaches to the gravitational self-force, there is nowhere any reference to a background geodesic. Instead, the residual fields govern the position of the puncture, and the position of the puncture effectively sources the residual fields.

To ensure that the metric perturbation is a solution to the Einstein equation, and not just the wave equations~\eqref{h1_eq}--\eqref{h2_eq}, we must ensure it satisfies the gauge condition~\eqref{gauge}. However, each $h^n_{\mu\nu}$ cannot satisfy a separate gauge condition of the form $\nabla^\nu\bar h^n_{\mu\nu}=0$, since such a condition is inconsistent with an accelerated worldline as a source. Instead, the perturbations together must satisfy $\e\nabla^\nu\bar h^1_{\mu\nu}+\e^2\nabla^\nu\bar h^2_{\mu\nu}=o(\e^2)$, with $\e\nabla^\nu\bar h^1_{\mu\nu}$ being on its own of order $\sim \e a^\mu$.\footnote{The precise condition on $h^1_{\mu\nu}$ can be written down explicitly. It is known~\cite{Pound:10a,Pound:12b} that the correct solution to Eq.~\eqref{h1_eq} is identical to that sourced by the point-mass stress-energy of Eq.~\eqref{T1_SC}, meaning $\bar h_1^{\mu\nu}(x)=4m\int_\gamma G^{\mu\nu}{}_{\mu'\nu'}(x,z(\tau))u^{\mu'}u^{\nu'}d\tau$, where a primed index indicates a tensor evaluated at $x'=z(\tau)$, and $G^{\mu\nu}{}_{\mu'\nu'}$ is a Green's function for the operator $E_{\mu\nu}$. Using the identity $G^{\mu\nu}{}_{\mu'\nu';\nu}=-G^\mu_{(\mu';\nu')}$, where $G^\mu{}_{\mu'}$ is a Green's function for the operator $\Box$ as it acts on a vector field, one finds the exact gauge condition to be $\nabla_{\!\nu}\bar h^{\mu\nu}=4m\int_\gamma G^\mu{}_{\mu'}\frac{D^2 z^{\mu'}}{d\tau^2}d\tau$.} In principle, these conditions should be satisfied automatically if the initial data satisfies it; one can verify this by taking the divergence of the first- and second-order wave equations and making use of the second-order Bianchi identity. In practice, however, gauge violations will be introduced numerically. Eliminating those violations should be possible with the introduction of constraint-damping terms~\cite{Barack-Lousto:05,Dolan-Barack:13}; for example, constraints of the form $0=Z^n_\mu\equiv \e^n\nabla^\nu\bar h^n_{\mu\nu} - \e^3 f^n_\mu$ might be used, where $n=1,2$ and $f^n_\mu$ are chosen vector fields that are uniformly of order 1. Constraints of this form do not affect the fields at orders $\e$ and $\e^2$, allowing them to maintain their correct relationship with the acceleration and thereby ensuring that the Einstein equation is satisfied through order $\e^2$.

\subsection{Gralla-Wald-type puncture scheme}
In a Gralla-Wald-type expansion of the field equations, rather than seeking an equation of motion for a self-accelerated worldline $\gamma$, one expands that worldline in a power series around a zeroth-order reference geodesic $\gamma_0$: given a coordinate description $z^\mu(s,\e)$ of $\gamma$, the expansion reads
\begin{equation}\label{z_expansion}
z^\mu(s,\e) = z_0^\mu(s)+\e z_1^\mu(s)+\e^2 z_2^\mu(s)+O(\e^3),
\end{equation}
where $s$ is a monotonic parameter along both $\gamma$ and $\gamma_0$. The leading-order term, $z_0^\mu(s)$, is the coordinate description of a geodesic of the background metric $g_{\mu\nu}$. The first-order term, $z_1^\mu\equiv \frac{\partial z^\mu}{\partial\e}|_{\e=0}$, is a vector on $\gamma_0$, describing the leading-order deviation of $\gamma$ from $\gamma_0$. The second-order term, if defined as $z_2^\mu\equiv \frac{1}{2}\frac{\partial^2z^\mu}{\partial^2\e}|_{\e=0}$, is simply a set of four scalars that depend on the choice of coordinates; because it is a second derivative (along a curve of increasing $\e$ and constant $s$), it does not transform as a vector.

A puncture scheme for this type of expansion can be derived from scratch in any gauge of choice, such as in Gralla's `P-smooth' gauges~\cite{Gralla:12}. Alternatively, a puncture scheme in the Lorenz gauge can be deduced simply by substituting the expansion~\eqref{z_expansion} into the metric perturbation~\eqref{h_SC_expansion}, the field equations~\eqref{h1_SC}--\eqref{h2_SC}, and the equation of motion~\eqref{motion_SC}, and then reorganizing terms according to explicit powers of $\e$. The metric perturbation is then given by the expansion
\begin{equation}\label{h_GW_expansion}
h_{\mu\nu} = \e h^1_{\mu\nu}[\gamma_0] + \e^2 h^2_{\mu\nu}[\gamma_0,z_1] + o(\epsilon^2).
\end{equation}
Here $h^1_{\mu\nu}$ is the same functional as in in Eq.~\eqref{h_SC_expansion}, but $\gamma_0$ has replaced $\gamma$ in its argument. On the other hand, $h^2_{\mu\nu}$ is now a different functional, which depends on $z_1$. Analogously, the decomposition into singular and regular fields in this expansion reads
\begin{align}
h^1_{\mu\nu} &= h^{\S1}_{\mu\nu}[\gamma_0] + h^{\R1}_{\mu\nu}[\gamma_0]= h^{\P1}_{\mu\nu}[\gamma_0] + h^{\res1}_{\mu\nu}[\gamma_0],\\
h^2_{\mu\nu} &= h^{\S2}_{\mu\nu}[\gamma_0,z_1] + h^{\R2}_{\mu\nu}[\gamma_0,z_1]= h^{\P2}_{\mu\nu}[\gamma_0,z_1] + h^{\res2}_{\mu\nu}[\gamma_0,z_1].
\end{align}

Near the object, the singular field takes the form
\begin{equation}\label{hS1_GW_schematic}
h^{\S1}_{\mu\nu}\sim \frac{m}{|x^i-z^i_0|} + O(|x^i-z^i_0|^0),
\end{equation}
\begin{equation}\label{hS2_GW_schematic}
h^{\S2}_{\mu\nu}\sim \frac{m^2+mz^\mu_{1\perp}}{|x^i-z^i_0|^2} + \frac{\delta m_{\mu\nu}+mh^{\res1}_{\mu\nu}}{|x^i-z^i_0|}
 + O(|x^i-z^i_0|^0).
\end{equation}
This form is identical to Eqs.~\eqref{hS1_SC_schematic}--\eqref{hS2_SC_schematic} but for two alterations:
\begin{itemize}
\item The divergent terms diverge on $\gamma_0$, not on $\gamma$.
\item The second-order singular field depends on the correction $z^\mu_1$ to the position.
\end{itemize}
The explicit expressions for the first few terms in these expansions, derived in Ref.~\cite{Pound:10a}, are given in Eqs.~\eqref{hS1_GW_Fermi} and \eqref{hS2_GW_Fermi}--\eqref{dm_GW_Fermi} in a local coordinate system $(t,x^i)$ centered on $\gamma_0$ (such that $z_0^i\equiv0$ in the schematic expressions above).

Because the point at which the puncture diverges is independent of the field values in this expansion, the puncture scheme becomes a sequence of equations, rather than a coupled system: first, the zeroth-order worldline is prescribed as a solution to the background geodesic equation,
\begin{equation}
\frac{D^2z^\mu_0}{d\tau_0^2} = 0,
\end{equation}
then the first order field is found from
\begin{subequations}\label{h1_GW}
\begin{align}
E_{\mu\nu}[h^{\res1}] &= -16\pi \bar T^1_{\mu\nu}[\gamma_0]-E_{\mu\nu}[h^{\P1}_{\alpha\beta}] \hspace{-5pt}& \text{inside }\Gamma_0,\\
E_{\mu\nu}[h^{1}] &= 0 & \text{outside } \Gamma_0,
\end{align}
\end{subequations}
then that field is used to find the first-order correction to the position by solving the Gralla-Wald equation~\cite{Gralla-Wald:08}
\begin{align}\label{motion_GW}
\frac{D^2z_{1\perp}^\mu}{d\tau_0^2} &= R^\mu{}_{\alpha\beta\gamma}u_0^\alpha u_0^\beta z_{1\perp}^\gamma \nonumber\\
			&\quad -\frac{1}{2}P_0^{\mu\gamma}\left(2h^{\res1}_{\gamma\alpha;\beta}-h^{\res1}_{\alpha\beta;\gamma}\right)u^\alpha_0 u^\beta_0,
\end{align}
and finally the second-order field is found from
\begin{subequations}\label{h2_GW}
\begin{align}
E_{\mu\nu}[h^{\res2}] &=2\delta^2R_{\mu\nu}[h^1,h^1]-16\pi \bar T^2_{\mu\nu}[\gamma_0,z_1]\hspace{-40pt}&\nonumber\\
											&\quad  - E_{\mu\nu}[h^{\P2}] & \text{inside }\Gamma_0,\\
E_{\mu\nu}[h^2] &= 2\delta^2R_{\mu\nu}[h^1,h^1] & \text{outside }\Gamma_0.
\end{align}
\end{subequations}
Here
\begin{align}
\bar T^1_{\mu\nu}[\gamma_0] &= \int_{\gamma_0} m(\tfrac{1}{2}g_{\mu\nu}+u_{0\mu} u_{0\nu}) \delta^4(x,z_0)d\tau_0,\label{T1_GW}\\
\bar T^2_{\mu\nu}[\gamma_0,z_1] &= \int_{\gamma_0} m(\tfrac{1}{2}g_{\mu\nu}+u_{0\mu} u_{0\nu})
														 z^{\gamma}_{1\perp}\frac{\partial}{\partial z_0^{\gamma}}\delta^4(x,z_0)d\tau_0\nonumber\\
													&\quad +\frac{1}{4}\int_{\gamma_0} \delta m_{\mu\nu}\delta^4(x,z_0)d\tau_0\label{T2_GW}
\end{align}
act as effective stress-energies sourcing the $m$, $\delta m_{\mu\nu}$, and $mz^a_1$ terms in Eqs.~\eqref{hS1_GW_schematic} and \eqref{hS2_GW_schematic}. $\tau_0$ is the proper time (measured in $g_{\mu\nu}$) on $\gamma_0$, $u_0^\mu\equiv \frac{dz_0^\mu}{d\tau_0}$ is the four-velocity on $\gamma_0$, $\frac{D}{d\tau_0}\equiv u_0^\mu\nabla_{\!\mu}$ is a covariant derivative along $u_0^\mu$,
\begin{equation}
P_0^{\mu\nu}\equiv g^{\mu\nu}+u_0^\mu u_0^\nu
\end{equation}
projects orthogonally to $u_0^\mu$, and
\begin{equation}
z^\mu_{1\perp}\equiv P_{0\nu}^{\mu} z_1^\nu
\end{equation}
is the piece of $z_1^\mu$ perpendicular to $u_0^\mu$.\footnote{Appendix D of Ref.~\cite{Pound:14a} describes why only the perpendicular piece of $z^\mu_{1}$ is needed as input for the second-order field. See also Sec.~\ref{GW_Fermi} below.} We have renamed the worldtube $\Gamma_0$ to indicate that $\gamma_0$ is always in its interior but $\gamma$ need not be.

If one wishes, as a final step in this procedure one can use the second-order field obtained from Eq.~\eqref{h2_GW} to find the second-order correction to the position; because that correction is not vectorial, we omit it, but we refer the reader to Refs.~\cite{Gralla:12,Pound:14a} for the differential equations governing $z_2^\mu$ (defined in particular local coordinate systems). 

In a scheme of this type, the correction to the motion is never incorporated into the position of the puncture, which diverges on the geodesic $z^\mu_0$ at all orders. This points to the fact that a Gralla-Wald-type expansion is valid only on timescales of order $\epsilon^0$, which are much much shorter than an inspiral time. After sufficient time, the correction $z_1^\mu$ will become large---as, for example, the small object falls into the large black hole in an EMRI---the series expansion of $z^\mu$ will no longer be valid, and the entire approximation scheme will fail. Nevertheless, a puncture scheme of this type can be useful for extracting short-term information about an inspiral, such as the conservative effects of the self-force at a given time~\cite{Pound:14c}. Because it is much easier to implement than a self-consistent scheme, it will likely be the preferred method for such calculations.

Unlike in the self-consistent expansion, here each of the perturbations must independently satisfy the Lorenz gauge condition:
\begin{align}
\nabla^\nu\bar h^1_{\mu\nu} =0=\nabla^\nu\bar h^2_{\mu\nu}.
\end{align}

\subsection{Building a practical puncture}
Several versions of the second-order self-field (and therefore several punctures) are now available. In Refs.~\cite{Pound:10a,Pound:12a,Pound:12b} one of us derived expressions for the self-field in the Lorenz gauge in both self-consistent and Gralla-Wald form in an arbitrary vacuum background and for an arbitrarily structured (sufficiently compact) small object. The last of these, Ref.~\cite{Pound:12b}, showed how the same can be done at arbitrary order in $\e$ in a broad class of wave gauges. In Ref.~\cite{Gralla:12}, Gralla presented a puncture scheme within a Gralla-Wald-type expansion in a broad class of P-smooth gauges in an arbitrary vacuum background for a nearly spherical and non-spinning object. 

However, all of these results were derived in local coordinate systems centered on the object's worldline (either $\gamma$ or $\gamma_0$). As of yet, no punctures have been presented in coordinate systems useful for numerical implementations of a puncture scheme. The main purpose of this paper is to fill that gap in the literature by deriving a covariant expansion of the second-order singular/self-field. From that covariant expansion, a puncture can be found in any desired coordinate system. We work in the Lorenz gauge, use the singular-regular split defined in Refs.~\cite{Pound:10a,Pound:12a}, and set the object's leading-order spin to zero. Our results are valid in any vacuum background.

We begin in Sec.~\ref{Fermi-field} with the field in Fermi-Walker coordinates $(t,x^i)$ centered on either $\gamma$ or $\gamma_0$. In these coordinates, the components of the singular field, through the orders that have been calculated, take the form
\begin{align}
h^{\S 1}_{\mu\nu}&=\sum_{p=-1}^2\sum_{\substack{\ell=0\\\ell\neq p}}^{p+1}r^p h^{(1p0\ell)}_{\mu\nu L}(t)\nhat^L+O(r^3),\\
h^{\S 2}_{\mu\nu}&=\sum_{p=-2}^1\sum_{\substack{\ell=0\\\ell\neq p}}^{p+4}r^p h^{(2p0\ell)}_{\mu\nu L}(t)\nhat^L\nonumber\\
								&\quad +(\ln r) \sum_{p=0,1}\sum_{\ell=p} r^p h^{(2p1\ell)}_{\mu\nu L}(t)\nhat^L\nonumber\\
								&\quad +O(r^2\ln r),\label{hS2_form_Fermi}
\end{align}
%\begin{align}
%h^{\S 1}_{\mu\nu}&=\sum_{p\geq-1}\sum_{\ell,q}r^p(\ln r)^q h^{\S(1pq\ell)}_{\mu\nu L}(t)\nhat^L,\\
%									&= \frac{2m\delta_{\mu\nu}}{r}+h^{\S(1001)}_{\mu\nu i}n^i
%											+r\left(h^{\S(1100)}_{\mu\nu}+h^{\S(1102)}_{\mu\nu ij}\nhat^{ij}\right)\nonumber\\
%									&\quad +r^2\left(h^{\S(1201)}_{\mu\nu i}n^i+h^{\S(1203)}_{\mu\nu ijk}\nhat^{ijk}\right)+O(r^3),\label{hS1_form}\\
%h^{\S 2}_{\mu\nu}&=\sum_{p\geq-2}\sum_{\ell,q}r^p(\ln r)^q h^{\S(2pq\ell)}_{\mu\nu L}(t)\nhat^L,\\
%									&= \frac{1}{r^2}\sum_{\ell=0,1,2}h^{\S(2,-2,0,\ell)}_{\mu\nu L}\nhat^L
%											+\frac{1}{r}\sum_{\ell=0,1,2,3}h^{\S(2,-1,0,\ell)}_{\mu\nu L}\nhat^L\nonumber\\
%									&\quad +(\ln r)h^{\S(2,0,1,0)}_{\mu\nu} +\sum_{\ell=1,2,3,4}h^{\S(2,0,0,\ell)}_{\mu\nu L}\nhat^L\nonumber\\
%									&\quad +r(\ln r)h^{\S(2,1,1,1)}_{\mu\nu i}n^i +r\sum_{\ell=0,2,3,4,5}h^{\S(2,1,0,\ell)}_{\mu\nu L}\nhat^L\nonumber\\
%									&\quad +O(r^2\ln r),\label{hS2_form}
%\end{align}
where $r\equiv\sqrt{\delta_{ab}x^ix^j}$ is the geodesic distance from the worldline. The first few of these terms are given explicitly in Eqs.~\eqref{hS1_SC_Fermi} and \eqref{hS2_SC_Fermi}--\eqref{hdm_SC_Fermi} in the self-consistent case and in Eqs.~\eqref{hS1_GW_Fermi} and \eqref{hS2_GW_Fermi}--\eqref{dm_GW_Fermi} in the Gralla-Wald case. All the terms displayed in Eq.~\eqref{hS2_form_Fermi} (i.e., terms through order $r$) were previously made available online~\cite{results}. In these expressions $n^i\equiv x^i/r$ is a unit vector pointing radially outward from the worldline, $L\equiv i_1\cdots i_\ell$ is a multi-index, and $\hat n^L\equiv n^{\langle i_1}\cdots n^{i_\ell\rangle}$ is the symmetric-trace-free (STF) product of $\ell$ unit vectors, with the trace defined with respect to $\delta_{ab}$. 

In Sec.~\ref{field-cov}, we put this in covariant form using the tools of near-coincidence expansions. The expansion parameter, in place of $r$, becomes $\sigma^{\mu'}\equiv \nabla^{\mu'}\sigma(x,x')$, where Synge's world function $\sigma(x,x')$ is equal to one-half the squared geodesic distance from $x'$ to $x$, the latter being the point off the worldline where the field is evaluated, and the former being an arbitrarily chosen nearby point on the worldline. We refer the reader to Ref.~\cite{Poisson-Pound-Vega:11} for a pedagogical introduction to covariant near-coincidence expansions and to Ref.~\cite{Heffernan-etal:12} for a recent example of their usage. The transformation from Fermi-Walker coordinates to covariant form is aided by the coordinates' convenient definition in terms of Synge's world function: $x^i\equiv-e^i_{\balpha}\sigma^{\balpha}$, where $e^i_\alpha$ is a triad leg on the worldline and a barred index signifies evaluation at a point $\bar x$ connected to $x$ by a geodesic intersecting the worldline orthogonally. Making use of this and similar definitions leads to, through the orders we have calculated, a covariant expansion of the form
\begin{align}
h^{\S 1}_{\mu\nu}&=g^{\mu'}_{\mu}g^{\nu'}_\nu \sum_{p=-1}^2\sum_{\ell=0}^{p+1}\sum_{i,j}\lambda^p\s^i\r^j \tilde h^{1p0ij\ell}_{\mu'\nu'\Lambda'}(x')\sigma^{\Lambda'}+O(\lambda^3),\\
h^{\S 2}_{\mu\nu}&=g^{\mu'}_{\mu}g^{\nu'}_\nu \Bigg[\sum_{p=-2}^1\sum_{\ell=0}^{p+4}\sum_{i,j}
					\lambda^p\s^i\r^j \tilde h^{2p0ij\ell}_{\mu'\nu'\Lambda'}(x')\sigma^{\Lambda'}\nonumber\\
					&\quad \ln(\lambda\s)\sum_{p=0}^1\sum_{\ell=0}^p\sum_{j=p-\ell}
					\lambda^p\r^j \tilde h^{2p10j\ell}_{\mu'\nu'\Lambda'}(x')\sigma^{\Lambda'}\Bigg]\nonumber\\
					&\quad +O(\lambda^2\ln\lambda),\label{hS2_form_cov}
%h^{\S 1}_{\mu\nu}&=g^{\mu'}_{\mu}g^{\nu'}_\nu\Big[\frac{2m(g_{\mu'\nu'}+2u_{\mu'}u_{\nu'})}{\s}K_{\mu'\nu'\alpha^\ell}\sigma^{}\\
%h^{\S 2}_{\mu\nu}&=g^{\mu'}_{\mu}g^{\nu'}_\nu\Big[
%					\lambda^{-2}\sum_{\ell=0}^2\sum_{\substack{p+q+\ell\\\ \ \  =-2}}\s^p\r^q K^{2,-2,p,q,\ell}_{\mu'\nu'\alpha^\ell}
%					\sigma^{\alpha^\ell}\nonumber\\
%					&\quad +\lambda^{-1}\sum_{\ell=0}^3\sum_{\substack{p+q+\ell\\\ \ \  =-2}}
%					\s^p\r^q K^{2,-1,pq\ell}_{\mu'\nu'\alpha^\ell}\sigma^{\alpha^\ell}\nonumber\\
%					&\quad +\lambda^{0}\sum_{\ell=0}^4\sum_{\substack{p+q+\ell\\\ \ \  =-2}}
%					\s^p\r^q K^{2pq\ell}_{\mu'\nu'\alpha^\ell}\sigma^{\alpha^\ell}\nonumber\\
%					&\quad +\lambda\sum_{\ell=0}^5\sum_{\substack{p+q+\ell\\\ \ \  =-2}}\s^p\r^q K^{2pq\ell}_{\mu'\nu'\alpha^\ell}\sigma^{\alpha^\ell}
%					+\ln(\lambda\s)[J^{2000}_{\mu'\nu'}+\s J^{2000}_{\mu'\nu'\alpha'}\sigma^{\alpha'}]+O(\lambda^2\ln\lambda)
\end{align}
where $\lambda$, introduced previously as a measure of spatial distance from the worldline, we now set equal to unity and use simply to count powers of that distance. The first few of these terms are given explicitly in Eqs.~\eqref{hS1_SC_cov}--\eqref{hdm_SC_final} in the self-consistent case and in Eqs.~\eqref{hS1_GW_cov}--\eqref{F1_GW} in the Gralla-Wald case. All the terms displayed in Eq.~\eqref{hS2_form_cov} (i.e., terms through order $\lambda$) are now available online~\cite{results}. In these expressions $g^{\mu'}_{\mu}$ is a parallel propagator from $x'$ to $x$, $\Lambda'\equiv\alpha_1'\cdots\alpha_\ell'$ is a multi-index, $\sigma^{\Lambda'}\equiv\sigma^{\alpha'_1}\cdots\sigma^{\alpha'_\ell}$, $\r$ and $\s$ are certain small distances defined in Eqs.~\eqref{r} and \eqref{s}, and the sums over $i$ and $j$ are such that $i+j+\ell=p$.

The covariant expansion of $h^{\S2}_{\mu\nu}$ represented by Eq.~\eqref{hS2_form_cov} is the central result of this paper. With that covariant expansion in hand, a puncture in any particular coordinate system can be easily found by expanding the covariant quantities $g^{\mu'}_\mu$ and $\sigma^{\mu'}$ in terms of coordinate distances $\Delta x^{\mu'}=x^\mu-x^{\mu'}$, where $x^{\alpha'}$ are the coordinate values at $x'$; see, e.g., Ref.~\cite{Heffernan-etal:12}. The result will be a puncture of the form
\begin{align}
h^{\P 1}_{\mu\nu}&=\sum_{p=-1}^2\delta^i_{2p+3}\delta^\ell_{3p+3}\frac{\lambda^p}{\rho^i}\mathcal{H}^{1p0i\ell}_{\mu'\nu'\Lambda'}(x')\Delta x^{\Lambda'},\\
h^{\P 2}_{\mu\nu}&=\sum_{p=-2}^1\ \sum_{\substack{i=2p+3\\i>0}}^{2p+8}\delta^\ell_{p+i}
								\frac{\lambda^p}{\rho^i}\mathcal{H}^{2p0i\ell}_{\mu'\nu'\Lambda'}(x')\Delta x^{\Lambda'}\nonumber\\
									&\quad +\ln(\lambda\rho) \sum_{p=0}^1\delta^\ell_p\lambda^p\mathcal{H}^{2p10\ell}_{\mu'\nu'\Lambda'}(x')\Delta x^{\Lambda'}.\label{hP2_coords}
\end{align}
where $\rho\equiv\sqrt{P_{\mu'\nu'}\Delta x^{\mu'}\Delta x^{\nu'}}$. A Gralla-Wald-type puncture $h^{\P 2}_{\mu\nu}$ of this form has already been calculated to order $\lambda\ln\lambda$ in the special case of circular orbits in Schwarschild coordinates~\cite{Pound:13b}. We leave the presentation of those and more general results to a future paper.

Within the body of the current paper, we display results of sufficiently high order to calculate the second-order regular field on the worldline. The results we present online~\cite{results} are of sufficiently high order to calculate both the second-order regular field on the worldline and the second-order force. We describe the precise order required of the puncture in our concluding discussion; readers uninterested in the technical details of our calculations may skip directly to that discussion. 

%Within using a puncture scheme within an $m$-mode (i.e., 2+1D), tensor-harmonic (i.e., 1+1D), or full spectral (i.e., 1D) decomposition. These results can be used in a Gralla-Wald-type scheme to calculate the second-order field but not the second-order correction to the position. They could also be used in a `first-order force, second-order field' self-consistent scheme, in which Eq.~\eqref{motion_SC} is replaced by the first-order equation of motion; such a scheme would still ensure that the second-order Einstein equation is satisfied through second order, albeit on shorter timescales than if Eq.~\eqref{motion_SC} is used. The results we present online~\cite{results} are of sufficiently high order to calculate both the second-order regular field on the worldline and the second-order force in a full 3+1D self-consistent puncture scheme. 

Although we have only derived results for the singular field of Refs.~\cite{Pound:10a,Pound:12a}, the same method could be used to generate a covariant expansion of Gralla's singular field~\cite{Gralla:12}, after first transforming from his choice of local coordinates to Fermi-Walker coordinates.

%%%%%%%%%%%%%%%%%%%%%%%%%%%%%%%%%%%%%%%%%%%%%%
\section{Singular field in Fermi-Walker coordinates}\label{Fermi-field}
%%%%%%%%%%%%%%%%%%%%%%%%%%%%%%%%%%%%%%%%%%%%%%
In both this and later sections, our strategy will be to discuss the self-consistent case first, working with the accelerated worldline. Afterward, we will state the results in the Gralla-Wald case by setting the acceleration of the worldline to zero and incorporating the position perturbation $z_1^\mu$. We first recapitulate the known results in Fermi-Walker coordinates.

\subsection{Background metric}
Fermi-Walker coordinates $(t,x^a)$ are constructed from a tetrad $(u^\alpha,e^\alpha_a)$ established along $\gamma$. The spatial triad is Fermi-Walker transported along the worldline according to
\begin{equation}\label{edot}
\frac{De^\alpha_a}{d\tau}=a_au^\alpha,
\end{equation}
where $a_a\equiv a_\mu e^\mu_a$ is a spatial component of $\gamma$'s acceleration, $a^\mu$. At each instant $\bar\tau$ of proper time, spatial geodesics are sent out orthogonally from the point $\bar x=z(\bar\tau)$ on $\gamma$. These geodesics generate a spatial hypersurface $\Sigma_{\bar\tau}$, and on that hypersurface coordinates $x^a$ are defined as
\begin{equation}\label{xa_def}
x^a=-e^a_{\balpha}\sigma^{\balpha}.
\end{equation}
The magnitude of these coordinates at a point $x$, given by $r\equiv \sqrt{\delta_{ab}x^ax^b}$, is the geodesic distance from $\bar x$ to $x$. $\sigma^{\balpha}$ is tangent to a generator of $\Sigma_{\bar\tau}$, satisfying
\begin{equation}\label{orthogonal}
\sigma_\balpha u^\balpha = 0.
\end{equation}
Each of the hypersurfaces is labelled with time $t=\bar\tau$, defining the coordinates $(t,x^a)$ at each point in the convex normal neighbourhood of $\gamma$.

Through order $r^3$, the metric in Fermi-Walker coordinates is given by
\begin{subequations}\label{background}%
\begin{align}
g_{tt} &= -1-2a_ix^i-\left(R_{0i0j}+a_ia_j\right)x^ix^j\nonumber\\
&\quad-\frac{1}{3}\left(4R_{0i0j}a_k+R_{0i0j|k}\right)x^ix^jx^k+O(r^4),\\
g_{ta} &= -\frac{2}{3}R_{0iaj}x^ix^j-\frac{1}{3}R_{0iaj}a_kx^ix^jx^k\nonumber\\
&\quad -\frac{1}{4}R_{0iaj|k}x^ix^jx^k+O(r^4),\\
g_{ab} &= \delta_{ab}-\frac{1}{3}R_{aibj}x^ix^j-\frac{1}{6}R_{aibj|k}x^ix^jx^k+O(r^4),
\end{align}
\end{subequations}
where the pieces of the Riemann tensor are evaluated on the worldline and contracted with members of the tetrad. For example, $R_{0iaj|k}\equiv R_{\balpha\bmu\bbeta\bnu;\bgamma} u^\balpha e^\bmu_ie^\bbeta_ae^\bnu_je^\bgamma_k$. An overdot will indicate a covariant derivative along the worldline, $\dot R_{0iaj}\equiv R_{\balpha\bmu\bbeta\bnu;\bgamma}u^\balpha e^\bmu_ie^\bbeta_ae^\bnu_ju^\bgamma$. We will later switch between this and the alternative notation $R_{\bar uiaj|k}\equiv R_{\balpha\bmu\bbeta\bnu;\bgamma} u^\balpha e^\bmu_ie^\bbeta_ae^\bnu_je^\bgamma_k$, as convenient.

Because the background is Ricci-flat, the components of the Riemann tensor and its first derivatives can be written in terms of Cartesian STF tensors $\E_{ab}$, $\B_{ab}$, $\E_{abc}$, and $\B_{abc}$, which we define as
\begin{align}
\E_{ab} &\equiv R_{0a0b}, \label{Eab}\\
\B_{ab} &\equiv \frac{1}{2}\epsilon^{pq}{}_{(a}R_{b)0pq}, \label{Bab}\\
\E_{abc} &\equiv \mathop{\rm STF}_{abc}R_{0a0b|c}, \label{Eabc}\\
\B_{abc} &\equiv \frac{3}{8}\mathop{\rm STF}_{abc}\epsilon^{pq}{}_{a}R_{b0pq|c},
\end{align}
where `STF' denotes the STF combination of the indicated indices. $\E_{ab}$ and $\B_{ab}$ are the even- and odd-parity tidal quadrupole moments of the background spacetime in the neighbourhood of $\gamma$, and $\E_{abc}$ and $\B_{abc}$ are the even- and odd-parity tidal octupole moments. Identities for decomposing each component of the Riemann tensor and its derivatives in terms of these tidal moments can be found in Appendix D3 of Ref.~\cite{Poisson-Vlasov:09}. We shall explicitly refer to the following of those identities:
\begin{align}
R_{abcd} &= \delta_{ac}\E_{bd}-\delta_{ad}\E_{bc}-\delta_{bc}\E_{ad}+\delta_{bd}\E_{ac},\label{Rabcd}\\
\E_{abc} &= \E_{ab|c} -\frac{1}{3}\left(\epsilon_{acp}\dot\B^p{}_b+\epsilon_{bcp}\dot\B^p{}_a\right),\\
\B_{abc} &= \frac{3}{4}\B_{ab|c}+\frac{1}{4}\left(\epsilon_{acp}\dot\E^p{}_b+\epsilon_{bcp}\dot\E^p{}_a\right).\label{Babc}
\end{align}

In addition to defining traces with respect to $\delta_{ab}$ when referring to tensors as STF, we raise and lower lowercase Latin indices with $\delta_{ab}$.

\subsection{Choice of singular and regular fields}\label{singular-regular}
The metric perturbation in Fermi-Walker coordinates has a local expansion~\cite{Pound:12b}
\begin{align}\label{h_ansatz}
h_{\mu\nu}^{n}&= \sum_{p\geq-n}\sum_{q,\ell}r^p(\ln r)^q h_{\mu\nu L}^{(npq\ell)}(t)\nhat^L,
\end{align}
with the cutoff at $p=-n$ following from the method of matched asymptotic expansions, in which we assume the existence of an `inner' expansion of the full metric around a background $g_{I\mu\nu}$ describing the geometry generated by the object if it were isolated. We define the $n$th-order singular-regular split of this field in terms of the coefficients $h_{\mu\nu L}^{(npq\ell)}$, which necessitates some preparatory discussion of them.

Substituting Eq.~\eqref{h_ansatz} into the wave equations~\eqref{h1_eq} and \eqref{h2_eq} transforms them into a sequence of Poisson equations of the form
\begin{equation}
\partial^a\partial_a\left[r^p(\ln r)^q h_{\mu\nu L}^{(npq\ell)}(t)\nhat^L\right] = P_{\mu\nu L}[h^{(n'<n,p'<p,q',L')}]\nhat^L,
\end{equation}
which can be solved order by order in $r$. As indicated, the source on the right-hand side depends on modes with lower $n$ and $p$. Since we begin with no source at the very lowest order ($n=1,\ p=-1$), it follows that when solving order by order in $r$, every mode $h_{\mu\nu L}^{(npq\ell)}$ will be written as a linear or nonlinear combination of the modes satisfying the homogeneous equation
\begin{equation}
\partial^a\partial_a\left[r^p h_{\mu\nu L}^{(np0\ell)}(t)\nhat^L\right] = 0.
\end{equation}
These special modes come in the forms
\begin{align}
\frac{1}{r^{\ell+1}}h_{\mu\nu L}^{(n,-\ell-1,0,\ell)}\nhat^L &\quad\text{for modes with } p<0,\\
r^\ell h_{\mu\nu L}^{(n,\ell,0,\ell)}\nhat^L &\quad\text{for modes with } p\geq0,
\end{align}
familiar from elementary electromagnetism. For more details, we refer the reader to Refs.~\cite{Pound:10a,Poisson-Pound-Vega:11,Pound:12b}.

The functions $h_{\mu\nu L}^{(n,-\ell-1,0,\ell)}(t)$ and $h_{\mu\nu L}^{(n,\ell,0,\ell)}(t)$ are determined by (i) the multipole moments of the spacetime $g_{I\mu\nu}$, (ii) the gauge condition, and (iii) global boundary conditions. Factor (i) relates the modes $h_{\mu\nu L}^{(n,-\ell-1,0,\ell)}$ to multipole moments of $g_{I\mu\nu}$ or corrections to them. Factor (ii) provides evolution equations for the multipole moments and relationships between the various modes. Our choice of singular-regular split is made in a way that is independent of global boundary conditions. Specifically, we define the regular field to be the piece of Eq.~\eqref{h_ansatz} containing no linear or nonlinear combinations of the modes $h_{\mu\nu L}^{(n,-\ell-1,0,\ell)}$; in other words, prior to imposing any global boundary conditions, it does not involve the object's multipole moments and is made up of freely specifiable functions. We define the singular field to be everything else in Eq.~\eqref{h_ansatz}, meaning $h^{\S n}_{\mu\nu}=h^{n}_{\mu\nu}-h^{\R n}_{\mu\nu}$.

With these definitions, the regular field possesses several nice properties~\cite{Pound:12a,Pound:12b,Pound:14a}:
\begin{itemize}
\item It is $C^\infty$ at $r=0$.
\item It is a solution to the vacuum Einstein equation; through second order that means $R_{\mu\nu}[g+h^\R]=O(\epsilon^3)$, including at $r=0$.
\item Through second order, the equation of motion is found to be equivalent to geodesic motion in the effective metric $g_{\mu\nu}+\e h^{\R1}_{\mu\nu}+\e^2 h^{\R2}_{\mu\nu}$ (assuming the object's leading-order spin and quadrupole moments are negligible). 
\end{itemize}
The singular field satisfies the following properties:
\begin{itemize}
\item In any domain that excludes $r=0$, its first- and second-order terms are solutions to the equations $\delta R_{\mu\nu}[h^{\S1}]=0$ and $E_{\mu\nu}[h^{\S2}]=2\delta^2R_{\mu\nu}[h^1,h^1]-2\delta^2R_{\mu\nu}[h^{\R1},h^{\R1}]$. If there exist boundary conditions for which $h^{\R1}_{\mu\nu}\equiv 0$, then with those boundary conditions and for $r\neq0$, $h^{\S}_{\mu\nu}$ satisfies the vacuum equation $R_{\mu\nu}[g+h^{\S}]=O(\epsilon^3)$.
\item In a domain including $r=0$, $h^{\S1}_{\mu\nu}$ is a solution to the wave equation with a point-mass source, $E_{\mu\nu}[h^{\S1}]=-8\pi m\delta_{\mu\nu}\delta^3(x^i)$, while $h^{\S2}_{\mu\nu}$ is not known to satisfy any distributionally well-defined equation. 
\item Unlike the regular field, it carries local information about the object's structure; it is made up entirely of terms that explicitly depend on the object's multipole moments or corrections to them.
\end{itemize}

One might suspect (or hope) that these lists of properties uniquely define the singular and regular fields. For example, one might like to think that the regular field can be defined to be the piece of the full field that is responsible for the self-force. This is untrue. Alternative choices could be made that satisfy all of the above properties; for example, we could split one of the functions $h^{(1,\ell,0,\ell)}_{\mu\nu L}$ with $\ell\geq2$ into two pieces, $h^{(1,\ell,0,\ell)}_{(1)\mu\nu L}$ and $h^{(1,\ell,0,\ell)}_{(2)\mu\nu L}$, and all terms in the solution~\eqref{h_ansatz} that are proportional to $h^{(1,\ell,0,\ell)}_{(2)\mu\nu L}$ could then be moved from the regular field to the singular field. However, of all possible ``nice" choices, ours arises most naturally in the process of solving the wave equations~\eqref{h1_eq} and \eqref{h2_eq} using the local expansion~\eqref{h_ansatz}: before making reference to any global boundary conditions, we simply put all the terms that involve the object's multipole moments into the singular field, and we put all the terms made up entirely of unknown functions into the regular field. At least through order $r^2$, the singular and regular fields $h^{\S1}_{\mu\nu}$ and $h^{\R1}_{\mu\nu}$ defined in this way coincide with those defined by Detweiler and Whiting~\cite{Detweiler-Whiting:03}. This can be seen concretely in the results displayed in Sec.~\ref{h1_SC_results} below.

We note one property the singular field does \emph{not} necessarily possess: it is not the case that every term $r^p(\ln r)^q h_{\mu\nu L}^{\S(npq\ell)}(t)\nhat^L$ in the singular field is of finite differentiability at $r=0$; the singular field, as we have defined it, can be expected to contain terms of that form that are $C^\infty$ functions of $x^i$. Using the explicit expression \eqref{hSS_SC_Fermi} below and Eq.~(32) of Ref.~\cite{Pound:12b}, we find that $h^{\S2}_{\mu\nu}$ will likely contain a term $r^2h_{tt}^{\S(2200)}(t)$, with $h_{tt}^{\S(2200)}\propto m^2\E_{ij}\E^{ij}$. However, at all orders we consider in the present paper, every term $r^p(\ln r)^q h_{\mu\nu L}^{\S(npq\ell)}(t)\nhat^L$ is of finite differentiability, and one can easily distinguish between contributions to the singular and regular fields using this fact.

We also note that the singular-regular split we use here (and which one of us used earlier in Refs.~\cite{Pound:10a,Pound:12a}) differs subtly from that used in Ref.~\cite{Pound:12b}. In the latter reference, the singular-regular split was defined in the identical way in terms of modes of the solutions to the wave equations, but for that paper's purposes, the modes referred to were those of the trace-reversed field. Appendix~\ref{trace-reverse} shows how the two definitions are related.

%\subsection{Self-consistent vs. Gralla-Wald expansions: mass dipole moment and acceleration}
\vspace{20pt}

\subsection{Self-consistent form}
\subsubsection{First order}
At first order, the singular field is completely described by the object's mass monopole $m=\frac{1}{2}h^{(1,-1,0,0)}_{tt}$, equal to the Arnowitt-Deser-Misner mass of the inner background metric $g_{I\mu\nu}$. The field is given by
\begin{widetext}
\begin{subequations}\label{hS1_SC_Fermi}
\begin{align}
h^{S1}_{tt} &= \frac{2m}{r}+3ma_i n^i+mr\left[4 a_{a} a^{a}+ (\tfrac{5}{3} \mathcal{E}_{ab} + \tfrac{3}{4} a_{a} a_{b}) 
								\hat{n}^{ab}\right]+mr^2\Big[\tfrac{9}{5} a_{a} a^{a} a_{b} \hat{n}^{b} 
								+ \tfrac{87}{20} \mathcal{E}_{ab} a^{a} \hat{n}^{b} \nonumber\\
							&\quad + \left(\tfrac{7}{12} \mathcal{E}_{abc} + \tfrac{3}{2} \mathcal{E}_{bc} a_{a} 
								- \tfrac{1}{8} a_{a} a_{b} a_{c}\right) \hat{n}^{abc}\Big]+O(r^3),\\
h^{S1}_{ta} &= mr\left[\tfrac{2}{3} \mathcal{B}^{bc} \epsilon_{acd} \hat{n}_{b}{}^{d} - 2 \dot a_{a}\right]
								+mr^2\Big[(\tfrac{7}{30} \mathcal{B}_{b}{}^{d} \epsilon_{acd} a^{b}
								- \tfrac{73}{30} \mathcal{B}_{c}{}^{d} \epsilon_{abd} a^{b}
								+ \tfrac{7}{6} \mathcal{B}_{a}{}^{d} \epsilon_{bcd} a^{b}) \hat{n}^{c}\nonumber\\
							&\quad	+ ( a_{b}\dot a_{a} - \tfrac{19}{30}  \dot{\mathcal{E}}_{ab} + 2  a_{a} \dot a_{b})\hat{n}^{b}   
								- \tfrac{1}{18}  \dot{\mathcal{E}}^{bc} \hat{n}_{abc} 
								+ \tfrac{7}{9}  \mathcal{B}^{cd} \epsilon_{ac}{}^{i} a^{b} \hat{n}_{bdi} 
								- \tfrac{2}{9} \mathcal{B}^{bcd} \epsilon_{ab}{}^{i} \hat{n}_{cdi}\Big]+O(r^3),\\
h^{S1}_{ab} &= \frac{2m\delta_{ab}}{r}-m\delta_{ab}a_in^i+mr\left[4 a_{a} a_{b}- \tfrac{38}{9} \mathcal{E}_{ab} 
								+ \tfrac{4}{3} \mathcal{E}_{(a}{}^{c} \hat{n}_{b)c} + \left(\tfrac{3}{4} a_{c} a_{d}
								- \mathcal{E}_{cd}\right)\delta_{ab} \hat{n}^{cd}\right]\nonumber\\
							&\quad +mr^2\Big[(\tfrac{11}{5} \mathcal{E}_{ab} a^{c} - 6 a_{a} a_{b} a^{c}) \hat{n}_{c}
								-\tfrac{2}{15}  \mathcal{E}_{c(a} a^{c} \hat{n}_{b)} 
								+ \tfrac{1}{12}  \mathcal{E}_{cd} a^{c} \delta_{ab} \hat{n}^{d} 
								+ (\tfrac{58}{15}  \mathcal{E}_{c(a} a_{b)}
								- \tfrac{31}{15}  \mathcal{E}_{abc} 
								-  \tfrac{68}{45}  \dot{\mathcal{B}}_{(a}{}^{d} \epsilon_{b)cd} 
								-  \tfrac{1}{4}  \delta_{ab} \ddot a_{c})\hat{n}^{c}\nonumber\\
							&\quad	+ \tfrac{2}{3}( \mathcal{E}_{(a}{}^{cd} 
								- \mathcal{E}_{(a}{}^{d} a^{c}) \hat{n}_{b)cd}  + \tfrac{2}{9}  \dot{\mathcal{B}}^{cd} \epsilon^i{}_{c(a} \hat{n}_{b)di} 
								+ (\tfrac{5}{6}  \mathcal{E}^{di} a^{c}- \tfrac{5}{12}  \mathcal{E}^{cdi}  
								-  \tfrac{5}{8}  a^{c} a^{d} a^{i}) \delta_{ab} \hat{n}_{cdi} \Big]+O(r^3),
\end{align}
\end{subequations}
\end{widetext}
where derivatives of the acceleration are defined as
\begin{align}
\dot a^\alpha \equiv \frac{Du^\alpha}{d\tau},\qquad \ddot a^\alpha \equiv \frac{D^2u^\alpha}{d\tau^2},
\end{align}
and
\begin{align}
\dot a^i \equiv e^i_{\bar\alpha}\dot a^{\bar\alpha} = \frac{da^i}{dt},\qquad
\ddot a^i \equiv e^i_{\bar\alpha}\ddot a^{\bar\alpha} = \frac{d^2a^i}{dt^2}.
\end{align}

This field satisfies $E_{\mu\nu}[h^{\S1}]=-8\pi m\delta_{\mu\nu}\delta^3(x^i)$. In covariant form, it satisfies 
\begin{equation}\label{hm_eq}
E_{\mu\nu}[h^{\S1}]=-16\pi \bar T^1_{\mu\nu}[\gamma],
\end{equation}
with $\bar T^1_{\mu\nu}[\gamma]$ given by Eq.~\eqref{T1_SC} above. In fact, Eq.~\eqref{hm_eq} serves to define $T^1_{\mu\nu}$; in this approach, the fact that the field is effectively sourced by a point-particle stress-energy tensor is a derived result, rather than an assumption.

The results for $h^{\S1}_{\mu\nu}$ in the particular form given here were first derived through order $r$ in Ref.~\cite{Pound:10a}. Their extension to order $r^2$ was reported in Ref.~\cite{Pound:12a} and given explicitly (in terms of the trace-reversed field) in Ref.~\cite{Pound:12b}.

\subsubsection{Second order}
At second order, the singular field is described by (i) the mass $m$, (ii) the first-order regular field, (iii) the correction $\delta m_{\mu\nu}\equiv h^{(2,-1,0,0)}_{\mu\nu}$ to the monopole moment, and (iv) the spin and mass dipole moments $S_i$ and $M_i$ of $g_{I\mu\nu}$, which make up $h^{(2,-2,0,1)}_{\mu\nu i}$. Throughout this paper, we set $S_i=0$. In the present case of a self-consistent expansion, we also set $M_i=0$, ensuring that the object is effectively mass-centered at the origin of our coordinates---which, recall, is $\gamma$. 

With those two terms set to zero, we may write the second-order singular field as the sum of three pieces:
\begin{equation}\label{hS2_SC_Fermi}
h^{\S2}_{\alpha\beta} = h^{\S\S}_{\alpha\beta}+h^{\S\R}_{\alpha\beta}+h^{\delta m}_{\alpha\beta}.
\end{equation}

The first piece,
\begin{subequations}\label{hSS_SC_Fermi}%
\allowdisplaybreaks\begin{align}
h^{\S\S}_{tt} &= -\frac{2m^2}{r^2}-\frac{10m^2a_in^i}{r} -m^2r^0\left(\tfrac{7}{3}\mathcal{E}_{ab} 
							+ \tfrac{29}{3} a_{a} a_{b}\right) \hat{n}^{ab}\nonumber\\
						&\quad+4m^2a_aa^a\ln r+O(r\ln r),\\
h^{\S\S}_{ta} &= m^2r^0\left(\tfrac{1}{2} \dot a_{b}\hat{n}_a{}^b - \tfrac{10}{3}\mathcal{B}^{bc} \epsilon_{acd} \hat{n}_{b}{}^{d}\right)-8m^2\dot a_a\ln r	\nonumber\\
						&\quad+O(r\ln r),\\
h^{\S\S}_{ab} &= \frac{\tfrac{8}{3}m^2\delta_{ab} - 7m^2\nhat_{ab}}{r^2}\nonumber\\
						&\quad+\frac{m^2}{r}\left[\tfrac{31}{5} a_{(a} n_{b)} 
							- \tfrac{37}{5} a^{c} \delta_{ab}n_{c} + \tfrac{14}{3}a^{c} \hat{n}_{abc}\right]\nonumber\\
						&\quad
							+ m^2r^0\Big[(4 \mathcal{E}_{c(a} - a_{c}a_{(a}) \hat{n}_{b)}{}^{c}- \tfrac{7}{2} a_{c} a^{c} \hat{n}_{ab}\nonumber\\
						&\quad + (\tfrac{10}{3} a_{c} a_{d}- \tfrac{4}{3} \mathcal{E}_{cd} )\delta_{ab} \hat{n}^{cd} + (\tfrac{7}{5}\mathcal{E}_{cd} 
							- \tfrac{56}{15} a_{c} a_{d})\hat{n}_{ab}{}^{cd}\Big]\nonumber\\
						&\quad+m^2\left(\tfrac{68}{15}a_ca^c\delta_{ab}-\tfrac{16}{15}\mathcal{E}_{ab}-\tfrac{8}{5}a_aa_b\right)\ln r\nonumber\\
						&\quad+O(r\ln r),
\end{align}
\end{subequations}
is a solution to $E_{\mu\nu}[h^{\S\S}]=2\delta^2 R_{\mu\nu}[h^{\S1},h^{\S1}]$ (off $r=0$). 

The second piece,
\begin{subequations}\label{hSR_SC_Fermi}%
\begin{align}
h^{\S\R}_{tt} &= -\frac{m h^{\R1}_{ab} \hat{n}^{ab}}{r}+O(r^0),\\
h^{\S\R}_{ta} &= -\frac{m h_{tb}^{\R1}\hat{n}_a{}^b}{r}+O(r^0),\\
h^{\S\R}_{ab} &= \frac{m}{r}\Big[2h^{\R1}_{c(a}\hat{n}_{b)}{}^c -\delta_{ab} h^{\R1}_{cd} \hat{n}^{cd} \nonumber\\
						&\quad - \left(h^{\R1}_{ij}\delta^{ij}+h_{tt}^{\R1}\right)\hat{n}_{ab}\Big] +O(r^0),
\end{align}
\end{subequations}
is a solution to $E_{\mu\nu}[h^{\S\R}]=2\delta^2 R_{\mu\nu}[h^{\S1},h^{\R1}]+2\delta^2 R_{\mu\nu}[h^{\R1},h^{\S1}]$  (off $r=0$). On the right-hand side of Eq.~\eqref{hSR_SC_Fermi}, the components of $h^{\R1}_{\mu\nu}$ are evaluated at $r=0$. At order $r^0$, $h^{\S\R}_{\mu\nu}$ also depends on first derivatives of $h^{\R1}_{\mu\nu}$ evaluated at $r=0$; at order $r$, it depends on second derivatives of $h^{\R1}_{\mu\nu}$ evaluated at $r=0$; and so on.

The final piece,
\begin{subequations}\label{hdm_SC_Fermi}
\begin{align}
h^{\delta m}_{tt} &= \frac{\delta m_{tt}}{r}+O(r^0),\\
h^{\delta m}_{ta} &= \frac{\delta m_{ta}}{r}+O(r^0),\\
h^{\delta m}_{ab} &= \frac{\delta m_{ab}}{r}+O(r^0),
\end{align}
\end{subequations}
is a solution to the homogeneous wave equation $E_{\mu\nu}[h^{\delta m}]=0$ off $r=0$. On a domain including $r=0$, it is a solution to the point-particle-like equation 
\begin{equation}
E_{\mu\nu}[h^{\delta m}] = -4\pi\delta m_{\mu\nu}(t)\delta^3(x^i),
\end{equation}
or, in covariant form,
\begin{equation}\label{hdm_eq}
E_{\mu\nu}[h^{\delta m}] = -16\pi \bar T^2_{\mu\nu}[\gamma],
\end{equation}
with $\bar T^2_{\mu\nu}[\gamma]$ given by Eq.~\eqref{T2_SC} above. As with $\bar T^1_{\mu\nu}$, this equation serves as a definition of $\bar T^2_{\mu\nu}$.

So far as the wave equation is concerned, each component of $\delta m_{\mu\nu}$ is an arbitrary function of time. The gauge condition constrains its components to be
\begin{subequations}\label{dm_SC_Fermi}
\begin{align}
\delta m_{tt} &= - \tfrac{1}{3} m h^{\R1}_{ab}[\gamma]\delta^{ab} - 2 m h_{tt}^{\R1}[\gamma],\\
\delta m_{ta} &= - \tfrac{4}{3} m h_{ta}^{\R1}[\gamma], \\
\delta m_{ab} &= \tfrac{2}{3} m h^{\R1}_{ab}[\gamma] + \tfrac{1}{3} m \delta_{ab} h^{\R1}_{cd}[\gamma]\delta^{cd}\nonumber\\
							&\quad + \tfrac{2}{3} m \delta_{ab} h_{tt}^{\R1}[\gamma],
\end{align}
\end{subequations}
up to an overall additive constant that we can write as $2\,\delta m\,\delta_{\mu\nu}$, which we choose to incorporate into $m$. Again, all components of the regular field in this expression are evaluated at $r=0$. We refer to $\delta m_{\mu\nu}$ as the correction to the object's monopole moment.

In the self-consistent scheme, it follows from the condition $M^i\equiv0$ and the gauge condition that the worldline $\gamma$ has acceleration
\begin{equation}\label{motion_SC_Fermi}
a_a = \frac{1}{2}\partial_a h^{\R1}_{tt}[\gamma]-\partial_t h^{\R1}_{ta}[\gamma]+O(\e^2).
\end{equation}

The results in this section were first derived  through order $r^0$ in Ref.~\cite{Pound:10a}, but with explicit appearances of $a^\mu$ set to zero. The extension to order $r$ and inclusion of acceleration terms were reported in Ref.~\cite{Pound:12a}. Reference~\cite{Pound:12b} presented results through order $r$ explicitly, but for a slightly different definition of the singular field, as mentioned above.

Because of the great length of the expressions in the expansions of $h^{\S2}_{\mu\nu}$ through order $r$, in this section we have begun our policy of only explicitly displaying the first few terms in the expansions; the complete results through order $r$ are available online~\cite{results}. In our concluding discussion in Sec.~\ref{required_order}, we explain the reasoning behind our choice of precisely \emph{which} orders to display in each of $h^{\S\S}_{\mu\nu}$, $h^{\S\R}_{\mu\nu}$, and $h^{\delta m}_{\mu\nu}$. To keep the present background section relatively concise, for the moment we say only that we display all terms that would be required to calculate the second-order regular field on the worldline using a practical numerical puncture scheme. All expansions must be written to one order higher to also calculate the second-order force in such a scheme.

\subsection{Gralla-Wald form}\label{GW_Fermi}
\subsubsection{First order}
The first-order singular field in a Gralla-Wald-type expansion is obtained from the self-consistent version by setting $a^\mu\equiv0$, leading to
\begin{equation}\label{hS1_GW_Fermi}
h^{\S1}_{\mu\nu}=\text{RHS of Eq.~\eqref{hS1_SC_Fermi} with $a^\mu\equiv 0$}.
\end{equation}
In this expression, $r$ now refers to the geodesic distance from $\gamma_0$ rather than $\gamma$. $h^{\S1}_{\mu\nu}$ satisfies Eq.~\eqref{hm_eq} with $\gamma\to\gamma_0$, $u^\mu\to u_0^\mu$, and $\tau\to\tau_0$.

\subsubsection{Second order}
The second-order singular field in a Gralla-Wald-type expansion is obtained from the self-consistent version by setting $a^\mu\equiv0$, making the replacement $\gamma\to\gamma_0$, and not setting the mass dipole moment $M_i$ to zero. We now identify $z_1^\mu$, the correction to the position, as $mz^i_1\equiv M^i$; or in covariant form,
\begin{equation}\label{M-z1}
m z^\alpha_{1\perp}\equiv M^\alpha
\end{equation}
(noting that $M^t$ appears nowhere in the metric, we set it to zero). That is, we take the value of the mass dipole moment of $g_{I\mu\nu}$ in Fermi coordinates centered on $\gamma_0$ to define the correction to the object's position relative to $\gamma_0$. The fact that only the perpendicular piece $z^\alpha_{1\perp}$ of this correction is relevant to the field will be discussed in Ref.~\cite{Pound:14a}. Intuitively, it is a consequence of the fact that we may always parametrize the family of worldlines $z^\mu(s,\lambda)$ in Eq.~\eqref{z_expansion} such that $z^\mu_1=z^\mu_{1\perp}$, which suggests that the piece of $z^\mu_1$ tangential to $u^\mu_0$ should have no real impact.

With the identification \eqref{M-z1} made, we may write the second-order singular field as
\begin{equation}\label{hS2_GW_Fermi}
h^{\S2}_{\alpha\beta} = h^{\S\S}_{\alpha\beta}+h^{\S\R}_{\alpha\beta}+h^{\delta m}_{\alpha\beta}+h^{\delta z}_{\alpha\beta}.
\end{equation}
$h^{\S\S}_{\alpha\beta}$ and $h^{\S\R}_{\alpha\beta}$ are given by Eqs.~\eqref{hSS_SC_Fermi} and \eqref{hSR_SC_Fermi}, with $a^\mu$ set to zero, $r$ now referring to geodesic distance from $\gamma_0$, and $h^{\R1}_{\mu\nu}[\gamma]$ replaced by $h^{\R1}_{\mu\nu}[\gamma_0]$. The  term $h^{\delta z}_{\mu\nu}$ is given by
\begin{subequations}\label{hdz_Fermi}
\begin{align}
h^{\delta z}_{tt} &= \frac{2mz_{1a}n^a}{r^2}+O(r^0),\\
h^{\delta z}_{ta} &= O(r^0),\\
h^{\delta z}_{ab} &= \frac{2mz_{1a}n^a}{r^2}+O(r^0).
\end{align}
\end{subequations}
$h^{\delta z}_{\mu\nu}$ is a solution to the homogeneous wave equation $E_{\mu\nu}[h^{\delta z}]=0$ off $r=0$. In a domain including $r=0$, it is a solution to the wave equation with a source equivalent to that created by the displacement of a point mass,
\begin{equation}
E_{\mu\nu}[h^{\delta z}] = 8\pi m\delta_{\mu\nu}z^a_1(t)\partial_a\delta^3(x^i),
\end{equation}
or in covariant form,
\begin{equation}\label{hdz_eq}
E_{\mu\nu}[h^{\delta z}] = -8\pi \int_{\gamma_0} m (g_{\mu\nu}+2u_{0\mu}u_{0\nu}) z^{\alpha}_{1\perp}\frac{\partial}{\partial z_0^\alpha}\delta^4(x,z)d\tau_0.
\end{equation}

The gauge condition now replaces the equation of motion~\eqref{motion_SC_Fermi} with the Gralla-Wald equation, given by 
\begin{equation}\label{motion_GW_Fermi}
\frac{d^2z_{1a}}{dt^2} = -\E_{ab}z^b_1+\frac{1}{2}\partial_a h^{\R1}_{tt}[\gamma_0]-\partial_t h^{\R1}_{ta}[\gamma_0],
\end{equation}
where the components of the regular field and its derivatives are evaluated at $r=0$.

$h^{\delta m}_{\mu\nu}$ here has the identical form as it did in the self-consistent case, and it continues to satisfy Eq.~\eqref{hdm_eq}(after making the replacements $\gamma\to\gamma_0$ and $\tau\to\tau_0$). But $\delta m_{\mu\nu}$ itself is modified by the effect of $z_1^a$:
\begin{subequations}\label{dm_GW_Fermi}
\begin{align}
\delta m_{tt} &= - \tfrac{1}{3} m h^{R1}_{ab}[\gamma_0]\delta^{ab} - 2 m h_{tt}^{R1}[\gamma_0],\\
\delta m_{ta} &= - \tfrac{4}{3} m h_{ta}^{R1}[\gamma_0] - 4m \dot z_{1a}, \\
\delta m_{ab} &= \tfrac{2}{3} m h^{R1}_{ab}[\gamma_0] + \tfrac{1}{3} m \delta_{ab} h^{R1}_{cd}[\gamma_0]\delta^{cd}\nonumber\\
							&\quad + \tfrac{2}{3} m \delta_{ab} h_{tt}^{R1}[\gamma_0].
\end{align}
\end{subequations}

The two fields $h^{\delta z}_{\mu\nu}$ and $h^{\delta m}_{\mu\nu}$ together satisfy
\begin{equation}
E_{\mu\nu}[h^{\delta z}+h^{\delta m}] = -16\pi \bar T^2_{\mu\nu}[\gamma_0,z_1],
\end{equation}
with $\bar T^2_{\mu\nu}[\gamma_0,z_1]$ given by Eq.~\eqref{T2_GW}. The $z^\mu_1$ terms in $T^2_{\mu\nu}$ are equivalent to those one would find by perturbing the worldline $\gamma$ in the point-mass stress-energy tensor~\eqref{T1_SC}; roughly speaking, the $z^\mu_1$ terms that appear by way of $h^{\delta m}_{\mu\nu}$ come from the correction to the four-velocity, and the terms that appear by way of $h^{\delta z}_{\mu\nu}$ come from expanding the argument of the delta function~\cite{Pound:14a}.

The results in this section were first derived through order $r^0$ in Ref.~\cite{Pound:10a}. The extension of $h^{\delta z}_{\mu\nu}$ to order $r$ is available online~\cite{results}.

%%%%%%%%%%%%%%%%%%%%%%%%%%%%%%%%%%%%%%%%%%%%%%%%%%%%%%
\section{Singular field in covariant form}\label{field-cov}
%%%%%%%%%%%%%%%%%%%%%%%%%%%%%%%%%%%%%%%%%%%%%%%%%%%%%%
\subsection{Outline of conversion strategy}
From the individual components $h^\S_{tt}$, $h^\S_{ta}$, and $h^\S_{ab}$, we now seek a covariant expansion of the tensor $h^\S$,
\begin{equation}\label{hS_tensor}
h^\S = h_{tt}^\S dt\otimes dt + h_{ta}^\S (dt\otimes dx^a+dx^a\otimes dt)+h_{ab}^\S dx^a\otimes dx^b.
\end{equation}
Here we have adopted index-free notation, and we have made tensor products explicit with a $\otimes$ to make clear that $dt$ and $dx^a$ cannot be commuted. 

We will derive our covariant expansion by expressing each quantity on the right-hand side of this equation in terms of Synge's world function. Rather than an expansion in powers of $r$, our final result will be an expansion in $\lambda$, which we remind the reader is used to count powers of spatial distance from the worldline on which $h^\S_{\mu\nu}$ diverges. We again work in the self-consistent case for the most part, with $h^\S_{\mu\nu}$ diverging on the accelerated worldline $\gamma$, and then afterward extract results in the Gralla-Wald case, with $h^\S_{\mu\nu}$ diverging on the zeroth-order worldline $\gamma_0$. 

Recall that the components are written as functions of $t$, $r$, and $n^a$. We will replace the dependence on $t$ with a dependence on $\bar x$, and we will replace $r$ and $n^a$ with
\begin{align}
r &= \sqrt{2\bar\sigma},\\
n^a &= \frac{-e^a_{\bar\alpha}\sigma^{\bar\alpha}}{\sqrt{2\bar\sigma}},
\end{align} 
which follow from the definition \eqref{xa_def}. In these expressions we have defined
\begin{equation}
\bar\sigma\equiv \sigma(x,\bar x).
\end{equation}
The one-forms $dt$ and $dx^a$ we will replace with
\begin{align}
dt &= \mu\sigma_{\bar\alpha\alpha}u^{\bar\alpha}dy^\alpha,\label{dt}\\
dx^a &= -e^a_{\bar\alpha}\left(\sigma^{\bar\alpha}{}_{\alpha}+\mu\sigma^{\bar\alpha}{}_{\bar\beta}u^{\bar\beta}\sigma_{\alpha\bar\gamma}u^{\bar\gamma}\right)dy^\alpha,\label{dxa}
\end{align}
where, e.g., $\sigma_{\bar\alpha\alpha}\equiv \sigma_{;\bar\alpha\alpha}$,
\begin{equation}
\mu = -\left(\sigma_{\bar\alpha\bar\beta}u^{\bar\alpha}u^{\bar\beta}+\sigma_{\bar\alpha}a^{\bar\alpha}\right)^{-1},
\end{equation}
and $y^\alpha$ are an arbitrary set of coordinates. These identities for the one-forms can be derived by taking total derivatives of Eqs.~\eqref{xa_def} and \eqref{orthogonal}~\cite{Poisson-Pound-Vega:11}.

After making these substitutions, we will have obtained the right-hand side of Eq.~\eqref{hS_tensor} in the form
\begin{multline}
h^\S_{\alpha\beta}dy^\alpha dy^\beta = \left[h_{tt}^\S \frac{\partial t}{\partial x^\alpha}\frac{\partial t}{\partial x^\beta} 
			+ 2 h_{ta}^\S \frac{\partial t}{\partial x^{(\alpha}}\frac{\partial x^a}{\partial x^{\beta)}}\right.\\
			 \left.+h_{ab}^\S \frac{\partial x^a}{\partial x^\alpha}\frac{\partial x^b}{\partial x^\beta}\right]dy^\alpha dy^\beta,
\end{multline}
with the quantity in square brackets written entirely in terms of tensors containing no remnant of Fermi-Walker coordinates. We will eliminate the dependence on the triad legs in this expression using the identity
\begin{equation}\label{ea.ea}
e^{\alpha}_a e^{a\beta} = P^{\alpha\beta}
\end{equation}
(or $e^{\alpha}_a e^{a\beta} = P_0^{\alpha\beta}$, in the Gralla-Wald case). We will then be left with a tensorial expression for $h^\S_{\alpha\beta}$ (now thinking of $\alpha$ and $\beta$ as abstract indices in the sense of Wald~\cite{Wald}).

\begin{figure}[t]
\begin{center}
\includegraphics[width=0.425\columnwidth]{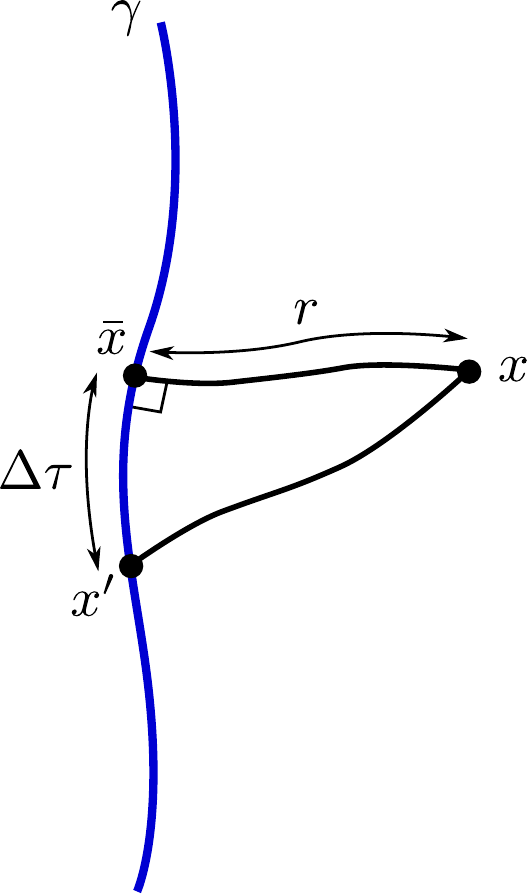}
\caption{\label{points} The point $x$ is connected to $\bar x$ by a unique geodesic that intersects $\gamma$ orthogonally. It is connected by a different unique geodesic to the point $x'$ on $\gamma$ that is separated from $\bar x$ by a proper time $\Delta\tau$.}
\end{center}
\end{figure}

For practical purposes, expressing the field at $x$ in terms of quantities at $\bar x$ is not ideal. $\bar x$ is always connected to $x$ by a geodesic that intersects $\gamma$ orthogonally, and if we wished to implement a puncture scheme in a particular coordinate system, we would have to express the coordinates at $\bar x$ in terms of the coordinates at $x$, which would create unnecessary complications. So rather than leaving our results in terms of $\bar x$, we expand the dependence on $\bar x$ about a nearby point $x'$ on $\gamma$. $x'$ is spatially related to $x$, but it is otherwise arbitrary. The general relationship between $x$, $\bar x$, and $x'$ is illustrated in Fig.~\ref{points}; since $x'$ is arbitrary, its specific relationship to $x$ can be chosen to maximize convenience. For example, $x$ and $x'$ can be made to have the same coordinate time in the coordinates one uses in one's numerics.

To express our quantities in terms of $x'$, we write $\bar x=z(\bar\tau)$ and $x'=z(\tau')$, and we expand in powers of 
\begin{equation}
\Delta\tau\equiv \bar\tau-\tau'.
\end{equation}
This procedure is made straightforward by the fact that each of the quantities $h^\S_{tt}$, $h^\S_{ta}$, $h^\S_{ab}$, $dt$, and $dx^a$ is a scalar at $\bar x$, meaning each can be expanded in an ordinary power series. So, for example, 
\begin{equation}
h^\S_{tt}=h^\S_{tt}(x,z(\bar\tau))=h^\S_{tt}(x,x')+\frac{dh^\S_{tt}}{d\tau'}(x,x')\Delta\tau+O(\Delta\tau).
\end{equation}

In the end, we wish our result to be in the form of a near-coincidence expansion in powers of $\sigma^{\alpha'}$. To achieve that, we will require the standard near-coincidence expansions~\cite{Christensen:76}
\begin{multline}
\sigma_{\alpha\beta'} = -g^{\alpha'}_\alpha\Big[g_{\alpha'\beta'} 
											+ \tfrac{1}{6} \lambda^2 R_{\alpha'\gamma'\beta'\zeta'} \sigma^{\gamma'} \sigma^{\zeta'} \\
											- \tfrac{1}{12} \lambda^3 R_{\alpha'\gamma'\beta'\zeta';\iota'}\sigma^{\gamma'}\sigma^{\zeta'}\sigma^{\iota'}
											+O(\lambda^4)\Big],
\end{multline}
\vspace{-1.5\baselineskip}
\begin{align}\label{sigmaBB}
\sigma_{\alpha'\beta'} &= g_{\alpha' \beta'} 
								- \tfrac{1}{3} \lambda^2 R_{\alpha'\gamma'\beta'\zeta'} \sigma^{\gamma'} \sigma^{\zeta'} \nonumber\\
								&\quad + \tfrac{1}{12} \lambda^3 R_{\alpha'\gamma'\beta'\zeta';\iota'}\sigma^{\gamma'} \sigma^{\zeta'} \sigma^{\iota'}
								+O(\lambda^4),
\end{align}
and
\begin{multline}\label{DBg}
g^{\alpha'}_{\mu;\mu'} = g^{\beta'}_{\mu}\Big[-\tfrac{1}{2} \lambda  R^{\alpha'}{}_{\!\!\beta'\mu'\gamma'} \sigma^{\gamma'} \\
									+ \tfrac{1}{6} \lambda^2 R^{\alpha'}{}_{\!\!\beta'\mu'\gamma';\zeta'} \sigma^{\gamma'}\sigma^{\zeta'}+O(\lambda^3)\Big].
\end{multline}
Expansions of higher derivatives of $\sigma$ and $g^{\mu'}_\mu$ can be generated recursively by taking derivatives of the above three equations.

As a check of our results, we can verify that each of the quantities $h^{\S1}_{\mu\nu}$, $h^{\S\S}_{\mu\nu}$, $h^{\S\R}_{\mu\nu}$, $h^{\delta m}_{\mu\nu}$, and $h^{\delta z}_{\mu\nu}$ is a solution to the appropriate field equation through the desired order; the field equation each should satisfy was described in the preceding section. For example, an expansion of $h^{\S1}_{\mu\nu}$ through order $\lambda^2$ should satisfy $E_{\mu\nu}[h^{\S1}]=0$ through order $\lambda^0$. To perform those checks, we make use of the standard near-coincidence expansion 
\begin{multline}
g^{\alpha'}_{\mu;\nu} = g^{\beta'}_{\mu}g^{\gamma'}_{\nu}\Big[-\tfrac{1}{2}\lambda R^{\alpha'}{}_{\!\!\beta'\gamma'\zeta'}\sigma^{\zeta'} \\
											+ \tfrac{1}{3} \lambda^2 R^{\alpha'}{}_{\!\!\beta'\gamma'\zeta';\iota'} \sigma^{\zeta'} \sigma^{\iota'}
											+O(\lambda^3)\Big].
\end{multline}

To facilitate notational simplicity of later expressions, we define the distances
\begin{equation}\label{r}
\r \equiv u_{\mu'}\sigma^{\mu'},
\end{equation}
which, in a rough sense, describes the proper time between $x'$ and $x$, and
\begin{equation}\label{s}
\s \equiv \sqrt{P_{\mu'\nu'}\sigma^{\mu'}\sigma^{\nu'}},
\end{equation}
which roughly describes the spatial distance between $x'$ and $x$. Both bits of notation are taken from Ref.~\cite{Haas-Poisson:06} by way of Ref.~\cite{Heffernan-etal:12}. In terms of these distances, we have the relation
\begin{equation}\label{s2-r2}
\sigma^{\mu'}\sigma_{\mu'} = 2\sigma(x,x') = \s^2-\r^2.
\end{equation}

\subsection{Orders of expansion}\label{strategy}
Before proceeding with the details, we summarize the strategy we will follow in presenting our results---specifically, the orders we carry the expansions to and the way we handle explicit dependence on acceleration.

A calculation of the second-order force requires $\partial h^{\res 2}_{\mu\nu}=\partial h^{\R 2}_{\mu\nu}$ on the worldline, meaning it requires $\partial h^{\P 2}_{\mu\nu}=\partial h^{\S 2}_{\mu\nu}+o(\lambda^0)$. This suggests that a puncture must satisfy $h^{\P 2}_{\mu\nu}=h^{\S 2}_{\mu\nu}+o(\lambda)$. Naively, we might infer that we must include four orders in $\lambda$ in our second-order puncture: all the terms from order $1/\lambda^2$ to order $\lambda$ (including the non-integer orders $\ln\lambda$ and $\lambda\ln\lambda$). Because $h^{\S\R}_{\mu\nu}$ at order $\lambda$ depends on second derivatives of $h^{\R1}_{\mu\nu}$ evaluated on the worldline, we must also include four orders in $\lambda$ in our first-order puncture: all the terms from order $1/\lambda$ to order $\lambda^2$. 

All of this would be true if a numerical puncture scheme were implemented in 3+1 dimensions. However, as we will explain in Sec.~\ref{required_order}, if the fields are decomposed into a suitable basis of functions such as azimuthal $m$-modes $e^{im\phi}$ or tensor spherical harmonics, the demands on the puncture are reduced, and some terms can be dropped. Taking this into account, we pursue the following strategy: to demonstrate our method, we display the full four orders in $\lambda$ in the expansions of $\Delta\tau$, $dt$, $dx^a$, and $h^{\S1}_{\mu\nu}$; but due to the length of the expansions of the various pieces of $h^{\S2}_{\mu\nu}$, for those pieces we display only the orders that would be required to calculate $h^{\R2}_{\mu\nu}$ on the worldline within an $m$-mode puncture scheme. Online, we present all four orders in $\lambda$~\cite{results}.

We follow a similar strategy in our treatment of explicit acceleration terms. For the expansions of $\Delta\tau$, $dt$, and $dx^a$, we treat the acceleration as arbitrary, such that the results can be used in whichever scheme one likes. However, in our expansions of the individual components of $h^\S_{\mu\nu}$, and in our final results for the covariant expansion of $h^\S_{\mu\nu}$, we take advantage of the fact that $a^\mu\sim\e$. We neglect all terms in $h^{\S2}_{\mu\nu}$ that explicitly depend on $a^\mu$, treating such terms as effectively higher order in $\e$. These neglected terms would alter the force only at $O(\e^3)$, and they need not be accounted for. Analogously, the only acceleration-dependent terms we include in $h^{\S1}_{\mu\nu}$ are those linear in $a^\mu$ and its derivatives. Such terms are effectively second order, and in practice, they can be transferred into the second-order puncture; this strategy is discussed in Sec.~\ref{order-reduction}. Because of this transfer to second order, we keep these linear-in-$a^\mu$ terms only through order $\lambda$, rather than $\lambda^2$.

\subsection{Expansion of $\Delta\tau$}
An expansion in powers of the proper time difference $\Delta\tau$ is useful only if we also possess a near-coincidence expansion of $\Delta\tau$ itself. We derive that expansion here.

In order to find $\Delta\tau$ in terms of Synge's world function, we define a function $p(\tau')=\sigma_{\alpha'}(x,z(\tau'))u^{\alpha'}$, and we expand $p(\bar\tau)$ around $p(\tau')$. Noting that $p(\bar\tau)=0$ [from Eq.~\eqref{orthogonal}], this yields
\begin{align}
0 &= \sigma_{\alpha'}u^{\alpha'} + \left(\sigma_{\alpha'\beta'}u^{\alpha'}u^{\beta'}+\sigma_{\alpha'}a^{\alpha'}\right)\Delta\tau \nonumber\\
	&\quad +\frac{1}{2}\left(\sigma_{\alpha'\beta'\gamma'}u^{\alpha'}u^{\beta'}u^{\gamma'} + 3\sigma_{\alpha'\beta'}u^{\alpha'}a^{\beta'}
		+\sigma_{\alpha'}\dot a^{\alpha'}\right)(\Delta\tau)^2 \nonumber\\
	&\quad +\frac{1}{6}\frac{D^3\sigma_{\alpha'}u^{\alpha'}}{d\tau'^3}(\Delta\tau)^3 + \frac{1}{24}\frac{D^4\sigma_{\alpha'}u^{\alpha'}}{d\tau'^4}(\Delta\tau)^4 + O[(\Delta\tau)^5].\label{sigma.u}
\end{align}
We have suppressed the explicit expressions for the third and fourth time derivatives, but they are easily obtained. We next expand derivatives of $\sigma_{\alpha'}$ near coincidence, using Eqs.~\eqref{sigmaBB} and \eqref{DBg}. To solve Eq.~\eqref{sigma.u} for $\Delta\tau$, we then expand $\Delta\tau$ in powers of $\lambda$ as
\begin{align}\label{Deltatau}
\Delta\tau &= \lambda\Delta_1\tau+\lambda^2\Delta_2\tau+\lambda^3\Delta_3\tau+\lambda^4\Delta_4\tau+O(\lambda^5)
\end{align}
and solve order by order in $\lambda$. After simplifying the expressions using the definitions~\eqref{s} and \eqref{r}, we find
\begin{subequations}
\begin{align}
\Delta_1 \tau &= \r, \\
\Delta_2 \tau &= \r a_{\sigma},\\
\Delta_3 \tau &= - \tfrac{1}{6} \r^3 a_{\alpha'} a^{\alpha'} + \tfrac{1}{2} \r^2 \dot{a}^{\sigma} + \r (a^{\sigma})^2
							-  \tfrac{1}{3} \r R_{u\sigma u\sigma},\\
\Delta_4 \tau &= - \tfrac{5}{24} \r^4 a^{\alpha'} \dot{a}_{\alpha'} + \tfrac{1}{6} \r^3\ddot{a}^{\sigma} 
							- \tfrac{2}{3} \r^3 a_{\alpha'} a^{\alpha'} a^{\sigma} + \tfrac{3}{2} \r^2 a^{\sigma} \dot{a}^{\sigma} \nonumber\\
							&\quad + \r (a^{\sigma})^3  -  \tfrac{1}{6} \r^3 a^{\alpha'}R_{\alpha' uu\sigma} 
							- \tfrac{1}{2} \r^2 a^{\alpha'}R_{\alpha'\sigma u\sigma} \nonumber\\
							&\quad - \tfrac{2}{3} \r a^{\sigma}R_{u\sigma u\sigma} - \tfrac{1}{8} \r^2 \dot R_{u \sigma u \sigma} 
							+ \tfrac{1}{12} \r R_{u \sigma u \sigma|\sigma}.
\end{align}
\end{subequations}

We have again borrowed notation from Ref.~\cite{Haas-Poisson:06} in defining, e.g., $R_{u \sigma u \sigma|\sigma}\equiv R_{\mu'\alpha'\nu'\beta';\gamma'}u^{\mu'}\sigma^{\alpha'}u^{\nu'}\sigma^{\gamma'}$. The contractions are always performed after taking derivatives along the worldline, such that, e.g., $\dot{a}^{\sigma}\equiv \dot{a}^{\mu'}\sigma_{\mu'}$.

\subsection{Expansion of $\sigma(x,\bar x)$}
Since the components $h^\S_{tt}$, $h^\S_{ta}$, and $h^\S_{ab}$ involve $r=\sqrt{2\bsigma}$, it is convenient to obtain an expansion of  $\sigma(x,\bar x)$ around $\sigma(x,x')$. We first expand in the interval of proper time,
\begin{align}
\sigma(x,z(\bar\tau)) &= \sigma(x,z(\tau'))+\frac{d\sigma}{d\tau'}\Delta\tau+\frac{1}{2}\frac{d^2\sigma}{d\tau'^2}\Delta\tau^2\nonumber\\
					&\quad +\frac{1}{3!}\frac{d^3\sigma}{d\tau'^3}\Delta\tau^3+\frac{1}{4!}\frac{d^4\sigma}{d\tau'^4}\Delta\tau^4\nonumber\\
					&\quad +\frac{1}{5!}\frac{d^5\sigma}{d\tau'^5}\Delta\tau^5+O(\lambda^6),
\end{align}
and we then substitute Eq.~\eqref{Deltatau} and the near-coincidence expansions of $\sigma_{\alpha'\beta'\cdots}$. The result is
\begin{align}\label{sigma_expansion}
\sigma(x,\bar x) &= \lambda^2\sigma_2(x,x')+\lambda^3\sigma_3(x,x')+\lambda^4\sigma_4(x,x')\nonumber\\
								&\quad +\lambda^5\sigma_5(x,x')+O(\lambda^6),
\end{align}
where, after simplifications involving Eqs.~\eqref{s}--\eqref{s2-r2}, the coefficients read
\begin{subequations}
\begin{align}
\sigma_2 &= \tfrac{1}{2} \s^2 \\
\sigma_3 &= \tfrac{1}{2} \r^2 a_{\sigma}\\
\sigma_4 &= \tfrac{1}{6} \r^3\dot{a}_{\sigma} - \tfrac{1}{24}\r^4 a_{\alpha'} a^{\alpha'}
						+ \tfrac{1}{2}\r^2 (a_{\sigma})^2 - \tfrac{1}{6}\r^2 R_{u\sigma u\sigma}\\
\sigma_5 &= - \tfrac{1}{24}\r^5 a^{\alpha'} \dot{a}_{\alpha'} + \tfrac{1}{24} \r^4\ddot{a}_{\sigma} 
						- \tfrac{1}{6}\r^4 a_{\alpha'} a^{\alpha'} a_{\sigma} + \tfrac{1}{2}\r^3 a_{\sigma} \dot{a}_{\sigma} \nonumber\\
					&\quad	+ \tfrac{1}{2}\r^2 (a_{\sigma})^3 - \tfrac{1}{24}\r^4 a^{\alpha'} R_{\alpha' uu\sigma} 
						- \tfrac{1}{6}\r^3 a^{\alpha'}R_{\alpha'\sigma u\sigma} \nonumber\\
					&\quad - \tfrac{1}{3} \r^2 a^{\sigma} R_{u\sigma u\sigma}
						- \tfrac{1}{24} \r^3 \dot R_{u \sigma u \sigma} + \tfrac{1}{24} \r^2 R_{u\sigma u\sigma|\sigma}.
\end{align}
\end{subequations}

\subsection{Expansions of $dt$ and $dx^a$}
The expansion of the one-forms $dt$ and $dx^a$ follows the same procedure as the expansion of $\sigma(x,\bar x)$: first expand in powers of $\Delta\tau$, then substitute Eq.~\eqref{Deltatau} and the near-coincidence expansion of derivatives of Synge's world function. In the case of $dx^a$, we will also have to make use of Eq.~\eqref{edot} for the derivative of $e^a_\alpha$ along the worldline.

It is helpful to first expand $\mu$ near coincidence; recall this quantity's appearance in Eqs.~\eqref{dt}--\eqref{dxa}. The result of that expansion is
\begin{align}
\mu &= 1 + \lambda a_{\bar\sigma} + \lambda^2 \left[(a_{\bar\sigma})^2 - \tfrac{1}{3} R_{\bar u \bar\sigma \bar u \bar\sigma}\right] \nonumber\\
 &\quad + \lambda^3 \left[(a_{\bar\sigma})^3 - \tfrac{2}{3} a_{\bar\sigma} R_{\bar u \bar\sigma \bar u \bar\sigma} + \tfrac{1}{12} R_{\bar u \bar\sigma \bar u \bar\sigma|\bar\sigma}\right]\nonumber\\
&\quad+O(\lambda^4),
\end{align}
where, e.g.,  $R_{\bar u \bar\sigma \bar u \bar\sigma}\equiv R_{\balpha\bbeta\bgamma\bdelta}u^\balpha \sigma^\bbeta u^\bgamma\sigma^\bdelta$. We place a bar over the subscripted $\sigma$'s and $u$'s to distinguish contracted quantities at $\bar x$ from those we defined at $x'$ as, e.g., $R_{u\sigma u\sigma}\equiv R_{\alpha'\beta'\gamma'\delta'}u^{\alpha'} \sigma^{\beta'} u^{\gamma'}\sigma^{\delta'}$.

Following the procedure in the case of $dt$, beginning from Eq.~\eqref{dt}, we arrive at
\begin{equation}\label{dt_expansion}
dt = \left[t_{0\mu}+\lambda t_{1\mu}+\lambda^2 t_{2\mu}+\lambda^3 t_{3\mu} +O(\lambda^4)\right]dx^\mu,
\end{equation}
where
\begin{subequations}
\begin{align}
t_{0\mu} &= - g^{\alpha'}_{\mu} u_{\alpha'},\\
t_{1\mu} &= - g^{\alpha'}_\mu \left(\r a_{\alpha'} + a_{\sigma}u_{\alpha'}\right),\\ 
t_{2\mu} &=	g^{\alpha'}_{\mu} \Big[\tfrac{1}{2} \r^2a_{\mu'} a^{\mu'} u_{\alpha'} - \r\dot{a}_{\sigma}u_{\alpha'} 
					- (a_{\sigma})^2 u_{\alpha'} - \tfrac{1}{2} \r^2\dot{a}_{\alpha'} \nonumber\\
				&\quad - 2\r a_{\alpha'} a_{\sigma} - \tfrac{2}{3} \r R_{\alpha'uu\sigma} - \tfrac{1}{6} R_{\alpha'\sigma u\sigma}
					 + \tfrac{1}{3} u_{\alpha'}R_{u\sigma u\sigma}\Big],\\
t_{3\mu} &= g^{\alpha'}_\mu\Big[\tfrac{2}{3} \r^3a_{\mu'} a^{\mu'} a_{\alpha'} -  \tfrac{3}{2} \r^2 a_{\sigma} \dot{a}_{\alpha'} 
					+ 2\r^2a_{\mu'} a^{\mu'} a_{\sigma}u_{\alpha'} \nonumber\\
				&\quad - 3 \r a_{\sigma} \dot{a}_{\sigma}u_{\alpha'} - (a_{\sigma})^3u_{\alpha'} 
					- \tfrac{1}{6}\r^3\ddot{a}_{\alpha'} - \tfrac{3}{2}\r^2 a_{\alpha'} \dot{a}_{\sigma} \nonumber\\
				&\quad - 3\r a_{\alpha'} (a_{\sigma})^2	+ \tfrac{2}{3}\r a_{\alpha'} R_{u\sigma u\sigma} 
					+ \tfrac{5}{6}\r^3a^{\mu'} \dot{a}_{\mu'} u_{\alpha'} \nonumber\\
				&\quad - \tfrac{1}{6} \r^3 a^{\mu'} R_{\alpha'u\mu' u} -  \tfrac{1}{2}\r^2 \ddot{a}_{\sigma} u_{\alpha'} 
					+ \tfrac{7}{12}\r^2 a^{\mu'} R_{\mu'\alpha'u\sigma}	\nonumber\\
				&\quad - \tfrac{1}{12}\r^2 a^{\mu'} R_{\mu' u\alpha'\sigma} 
					-  \tfrac{5}{12}\r^2 a^{\mu'} R_{\mu'\sigma\alpha'u}	-  \tfrac{1}{6} a^{\mu'} \r R_{\mu'\sigma\alpha'\sigma} \nonumber\\
				&\quad + \tfrac{1}{2} a^{\mu'} \r^2 u_{\alpha'} R_{\mu' uu\sigma} -  \tfrac{4}{3} \r a_{\sigma} R_{\alpha'uu\sigma} 
					+ a^{\mu'} \r u_{\alpha'} R_{\mu'\sigma u\sigma} \nonumber\\
				&\quad - \tfrac{1}{6} a_{\sigma} R_{\alpha'\sigma u\sigma} 
					+ \tfrac{2}{3} a_{\sigma} u_{\alpha'} R_{u\sigma u\sigma} - \tfrac{1}{4} \r^2 \dot R_{\alpha'uu \sigma}\nonumber\\
				&\quad - \tfrac{1}{12} \r\dot R_{\alpha' \sigma u \sigma} + \tfrac{1}{4} \r R_{\alpha' uu\sigma|\sigma} 
					+ \tfrac{1}{12}R_{\alpha'\sigma u\sigma|\sigma} \nonumber\\
				&\quad + \tfrac{1}{4} \r u_{\alpha'} \dot R_{u\sigma u\sigma} -  \tfrac{1}{12} u_{\alpha'} R_{u\sigma u\sigma|\sigma}\Big].
\end{align}
\end{subequations}

Following the procedure in the case of $dx^a$, beginning from Eq.~\eqref{dxa}, we arrive at
\begin{equation}\label{dx_expansion}
dx^a = \left[x^a_{0\mu}+\lambda x^a_{1\mu}+\lambda^2 x^a_{2\mu}+\lambda^3 x^a_{3\mu} +O(\lambda^4)\right]dx^\mu,
\end{equation}
where
\begin{subequations}
\begin{align}
x^a_{0\mu} &= g^{\beta'}_\mu e^a_{\beta'}, \\
x^a_{1\mu} &= g^{\beta'}_{\mu} e^{a\alpha'}\r u_{\beta'}a_{\alpha'},\\
x^a_{2\mu} &= g^{\beta'}_{\mu}e^{a\alpha'}\Bigl(\tfrac{1}{2}\r^2 a_{\alpha'} a_{\beta'} + \r a_{\alpha'} a_{\sigma} u_{\beta'} 
			+\tfrac{1}{2}\r^2 \dot{a}_{\alpha'} u_{\beta'} \nonumber\\
			&\quad + \tfrac{1}{6}\r^2 R_{\alpha'u\beta'u} - \tfrac{1}{2}\r R_{\alpha'\beta'u\sigma}
			-\tfrac{1}{3} \r R_{\sigma(\beta'\alpha')u}\nonumber\\
			&\quad + \tfrac{1}{6} R_{\alpha'\sigma\beta'\sigma} - \tfrac{1}{3} \r u_{\beta'} R_{\alpha'uu\sigma} 
			- \tfrac{1}{3} u_{\beta'}R_{\alpha'\sigma u\sigma}\Bigr), 
\end{align}
\begin{align}
x^a_{3\mu} &= g^{\beta'}_{\mu}e^{a\alpha'} \Bigl[\r a_{\alpha'} (a_{\sigma})^2 u_{\beta'} 
			+ \tfrac{1}{3}\r^3 a_{\beta'} \dot{a}_{\alpha'} + \tfrac{1}{6}\r^3 a_{\alpha'} \dot{a}_{\beta'}\nonumber\\
			&\quad + \r^2a_{\alpha'} a_{\beta'} a_{\sigma} -  \tfrac{1}{3}\r^2 a_{\beta'} R_{\alpha'uu\sigma} 
			- \tfrac{1}{3}\r a_{\beta'} R_{\alpha'\sigma u\sigma}\nonumber\\ 
			&\quad - \tfrac{1}{3} \r^3 a_{\alpha'} a_{\mu'} a^{\mu'} u_{\beta'} + \tfrac{1}{6} \r^3 a^{\mu'} R_{\alpha'[\mu'\beta']u}\nonumber\\
			&\quad - \tfrac{1}{6}\r^3 a^{\mu'}u_{\beta'} R_{\alpha'u\mu'u} + \tfrac{1}{12}\r^3 a^{\mu'} R_{\alpha'u\beta'\mu'} \nonumber\\
			&\quad + \r^2 a_{\sigma} \dot{a}_{\alpha'} u_{\beta'} + \tfrac{1}{2} \r^2 a_{\alpha'} \dot{a}_{\sigma} u_{\beta'}
			 + \tfrac{1}{3}\r^2 a_{\sigma} R_{\alpha'u\beta'u}\nonumber\\ 
			&\quad - \tfrac{1}{6}\r^2 a^{\mu'} R_{\mu'(\alpha'\beta')\sigma} - \tfrac{1}{4}\r^2 a^{\mu'} R_{\alpha'\beta'\mu'\sigma} \nonumber\\
			&\quad + \tfrac{1}{3}\r^2  a^{\mu'}u_{\beta'}R_{\alpha'[\mu'\sigma]u}
			 - \tfrac{1}{3} \r^2 a^{\mu'}u_{\beta'}R_{\alpha'u\mu'\sigma} \nonumber\\
			&\quad -  \tfrac{1}{2} \r a_{\sigma}R_{\alpha'\beta' u\sigma} 
			- \tfrac{1}{3} \r a_{\sigma} R_{u(\alpha'\beta')\sigma} \nonumber\\
			&\quad-  \tfrac{1}{3}\r a^{\mu'} u_{\beta'} R_{\alpha'\sigma\mu'\sigma}
 			-  \tfrac{2}{3}\r a_{\sigma} u_{\beta'} R_{\alpha'uu\sigma}\nonumber\\
			&\quad- \tfrac{1}{3} a_{\sigma} u_{\beta'} R_{\alpha'\sigma u\sigma} 
			+ \tfrac{1}{6}\r^3 \ddot{a}_{\alpha'} u_{\beta'} + \tfrac{2}{3} \r^2 a_{\alpha'} R_{\beta'uu\sigma} \nonumber\\
			&\quad + \tfrac{1}{6}\r a_{\alpha'} R_{\beta'\sigma u\sigma} + \tfrac{1}{12} \r^3 \dot R_{\alpha' u\beta'u} 
			- \tfrac{1}{6} \r^2 \dot R_{\alpha' \beta' u\sigma} \nonumber\\
			&\quad - \tfrac{1}{6} \r^2 \dot R_{\sigma(\beta'\alpha')u} 
			+ \tfrac{1}{12} \r \dot R_{\alpha' \sigma\beta' \sigma} + \tfrac{1}{6} \r R_{\alpha' \beta'u\sigma|\sigma}\nonumber\\
			&\quad - \tfrac{1}{12} \r^2 R_{\alpha'u\beta'u|\sigma} + \tfrac{1}{6} \r R_{\sigma(\beta'\alpha')u|\sigma} 
			- \tfrac{1}{12} R_{\alpha' \sigma\beta' \sigma|\sigma}\nonumber\\ 
			&\quad - \tfrac{2}{3} \r a_{\alpha'} u_{\beta'} R_{u\sigma u\sigma}
 			-  \tfrac{1}{4} \r^2 u_{\beta'} \dot R_{\alpha'uu\sigma} -  \tfrac{1}{4} \r u_{\beta'} \dot R_{\alpha' \sigma u\sigma} \nonumber\\
			&\quad + \tfrac{1}{12} \r u_{\beta'}R_{\alpha' uu\sigma|\sigma} 
			+ \tfrac{1}{12} u_{\beta'}R_{\alpha' \sigma u\sigma|\sigma}\Bigr].
\end{align}
\end{subequations}

\subsection{Self-consistent form}
\subsubsection{Expansions of $h^\S_{tt}$, $h^\S_{ta}$, and $h^\S_{ab}$}
We now turn to the covariant expansion of the individual components. For each component, we begin by expressing it in terms of covariant bitensorial quantities at $x$ and $\bar x$. We then follow the same procedure as above, expanding in powers of $\Delta\tau$ and following that with a near-coincidence expansion. While we have strived for generality in our expansions of $\Delta\tau$, $\bar\sigma$, $dt$, and $dx^a$, displaying four orders in $\lambda$ for each of them and keeping all acceleration terms, we now limit our inclusion of acceleration terms and reduce the number of orders that we explicitly display, following the reasoning outlined in Sec.~\ref{strategy}.

When rewriting the components in terms of covariant quantities, one of our tasks is to replace the STF 3-tensors $\nhat^L$ with 4-tensorial quantities evaluated at $\bar x$. As an example, 
\begin{align}
\nhat^{ab} &= n^an^b-\frac{1}{3}\delta^{ab}\\
					&= e^a_\balpha e^b_\bbeta \left(\frac{\sigma^\balpha\sigma^\bbeta}{2\bar\sigma}-\frac{1}{3}P^{\balpha\bbeta}\right),
\end{align}
where we have used Eq.~\eqref{ea.ea}.
We will also have to rewrite components of other 3-tensors such as $h^{\R1}_{ab}(t,0)$ in terms of 4-tensors, which we do as
\begin{equation}
h^{\R1}_{ab}(t,0)=h^{\R1}_{\balpha\bbeta}e^\balpha_a e^\bbeta_b.
\end{equation}
So, for example, in contractions of 3-tensors we arrive at expressions such as
\begin{equation}
h^{\R1}_{ab}\nhat^{ab} = h^{\R1}_{\balpha\bbeta}\left(\frac{\sigma^\balpha\sigma^\bbeta}{2\bar\sigma}-\frac{1}{3}P^{\balpha\bbeta}\right),
\end{equation}
where we have again used Eq.~\eqref{ea.ea} and appealed to the fact that $P^\balpha{}_\bbeta\sigma^\bbeta=\sigma^\balpha$. We shall also require conversions from the 3-tensors $\E_{ab}$, $\B_{ab}$, $\E_{abc}$, and $\B_{abc}$. The first two of these are trivially obtained from Eqs.~\eqref{Eab} and \eqref{Bab}. The latter two are given by 
\begin{align}
\E_{abc} &= \frac{1}{3}e^\balpha_a e^\bbeta_b e^\bgamma_c\left(R_{\balpha\bar u \bbeta\bar u;\bgamma}+R_{\bgamma\bar u \balpha\bar u;\bbeta}+R_{\bbeta\bar u \bgamma\bar u;\balpha}\right),\\
\epsilon_{abp}\B^p{}_{cd} &= \frac{1}{4}e^\balpha_a e^\bbeta_b e^\bgamma_c e^\bdelta_d\bigg[3R_{\balpha\bbeta\bgamma\bar u;\bdelta}-\dot R_{\balpha\bbeta\bgamma\bdelta}+2P_{\bdelta[\balpha}\dot R_{\bbeta]\bar u\bgamma\bar u}\bigg],
\end{align}
which follow from Eqs.~\eqref{Eabc} and \eqref{Babc}.

We first examine the components of the first-order field, given in Eq.~\eqref{hS1_SC_Fermi}. Written in terms of tensors at $\bar x$, they read
\begin{subequations}
\begin{align}
h^{\S1}_{tt} &= \frac{m}{\sqrt{2\bsigma}}\left(2\lambda^{-1}-3\lambda^0a_\bsigma
								+\tfrac{5}{3}\lambda R_{\bu\bsigma\bu\bsigma}-\tfrac{7}{12}\lambda^2R_{\bsigma\bu\bsigma\bu|\bsigma}\right)\nonumber\\
						 &\quad +O(\lambda a^2,\lambda^2 a,\lambda^3) \\
h^{\S1}_{ta} &= \frac{me^\balpha_a}{\sqrt{2\bsigma}}\left[\lambda\left(\tfrac{2}{3}R_{\balpha\bsigma\bsigma\bu}
								-4\bsigma\dot a_\balpha\right)+\lambda^2\left(\tfrac{4}{3}\bsigma \dot R_{\balpha\bu\bsigma\bu}\right.\right.\nonumber\\
							&\quad \left.\left.-\tfrac{1}{6}R_{\balpha\bsigma\bsigma\bu|\bsigma}\right)\right] +O(\lambda a^2,\lambda^2 a,\lambda^3),\\
h^{\S1}_{ab} &= \frac{me_a^\balpha e_b^\bbeta}{\sqrt{2\bsigma}}\left[2\lambda^{-1}g_{\balpha\bbeta} + \lambda^0 g_{\balpha\bbeta}a_{\bsigma}
							- \lambda \bigl(\tfrac{2}{3} R_{\balpha \bsigma\bbeta \bsigma}\right.\nonumber\\
						&\quad \left.+ \tfrac{1}{3}g_{\balpha\bbeta} R_{\bu\bsigma\bu\bsigma} + 8\bsigma R_{\balpha \bu \bbeta \bu}\right) 
						+ \lambda^2 \big(\tfrac{1}{12}g_{\balpha\bbeta} R_{\bu\bsigma \bu\bsigma|\bsigma}\nonumber\\
 						&\quad + \left.\tfrac{1}{3} R_{\balpha\bsigma \bbeta\bsigma|\bsigma} 
						+ 4\bsigma R_{\balpha\bu\bbeta\bu|\bsigma}\big)\right] +O(\lambda a^2,\lambda^2 a,\lambda^3).
\end{align}
\end{subequations}
We have simplified the $ab$ components using the identities
\begin{align}
2\E_{c(a}n_{b)} n^c &= \E_{ab}+\delta_{ab}\E_{cd}n^cn^d-R_{acbd}n^cn^d,\label{ENN}\\
2\E_{be(a}n_{c)} n^b &= -R_{abcd|e}n^dn^b+\delta_{ac}\left(\E_{bde}-\tfrac{2}{3}\dot R_{ebd0}\right)n^dn^b\nonumber\\
									&\quad +\E_{ace}-\tfrac{2}{3}\dot R_{e(ac)0}+\tfrac{2}{3}\dot R_{e(bc)0}n_an^b \nonumber\\
									&\quad -\left(\E_{ade}-\tfrac{2}{3}\dot R_{e(ad)0}\right)n^dn_c,
\end{align}
and
\begin{align}
R_{\balpha\bu\bsigma\bu|\bbeta} + R_{\bbeta\bu\bsigma\bu|\balpha} = 2\dot R_{\bsigma(\balpha\bbeta)\bu}+ 2R_{\balpha\bu\bbeta\bu|\bsigma},
\end{align}
which follow from Eqs.~\eqref{Rabcd}--\eqref{Babc}.

Expanding the dependence on $\bar x$ about $x'$, as, e.g., $h^{\S1}_{ab}(x,z(\bar\tau))=h^{\S1}_{ab}(x,z(\tau'))+\frac{d h^{\S1}_{ab}}{d\tau}(x,z(\tau'))\Delta\tau+\ldots$, and then using Eq.~\eqref{Deltatau}, Eq.~\eqref{sigma_expansion}, the near-coincidence expansions of derivatives of $\sigma_{\alpha'}$, and Eq.~\eqref{edot}, we find
\begin{subequations}\label{hS1_expansion}%
\begin{align}
h^{\S1}_{tt} &= \frac{2 m}{\lambda \s} - \frac{m\lambda^0}{\s^3}\left(\r^2+3\s^2\right) a_{\sigma}
							+\frac{m\lambda}{3\s^3} \left[\left(\r^2+5\s^2\right)R_{u\sigma u\sigma}\right.\nonumber\\
							&\quad  \left.- \left(\r^3+9\r\s^2\right)\dot{a}_\sigma\right] 
							-\frac{m \lambda^2}{12\s^5} \left(\r^2 \s^2R_{u\sigma u\sigma|\sigma}  - \r^3 \s^2 \dot R_{u\sigma u\sigma}\right.\nonumber\\
							&\quad \left.+7\s^4R_{u\sigma u\sigma|\sigma}-13\r\s^4\dot R_{u\sigma u\sigma}\right)+O(\lambda a^2,\lambda^2 a,\lambda^3),\\
h^{\S1}_{ta} &= -me_a^{\alpha'}\biggl\{\frac{2\lambda}{3\s}\left(R_{\alpha'\sigma u\sigma}- \r R_{\alpha'u\sigma u} 
								+ 3 \s^2\dot{a}_{\alpha'}\right)\nonumber\\ 
							&\quad + \frac{m \lambda^2}{18\s^3}\left[9\r\s^2\dot R_{\alpha'\sigma u\sigma}-3\s^2R_{\alpha'\sigma u\sigma|\sigma}
								+ 3\r\s^2R_{\alpha'u\sigma u|\sigma}\right.\nonumber\\
							&\quad \left.- \left(9\r^2\s^2+12\s^4\right)\dot R_{\alpha'u\sigma u}\right]\biggr\}
								+O(\lambda a^2,\lambda^2 a,\lambda^3),\\
h^{\S1}_{ab} &= me_{(a}^{\alpha'}e_{b)}^{\beta'}\bigg\{ \frac{2 g_{\alpha'\beta'}}{\lambda\s}
							+\frac{\lambda^0}{\s^3}g_{\alpha'\beta'}a_{\sigma} \left(\s^2-\r^2\right)\nonumber\\
							&\quad + \frac{\lambda}{3\s^3}\left[4\r\s^2R_{u(\alpha'\beta')\sigma}-2\s^2 R_{\alpha'\sigma\beta'\sigma} -12\s^4R_{\alpha'u\beta'u} \right.\nonumber\\
							&\quad  \left. +g_{\alpha'\beta'}(\r^2-\s^2)R_{u\sigma u\sigma} 
							- g_{\alpha'\beta'}\r(\r^2-3\s^2)\dot{a}_\sigma)\right]\nonumber\\
							&\quad + \frac{\lambda^2}{12\s^3}\left[g_{\alpha'\beta'}(\r^3- 3\r\s^2)\dot R_{u\sigma u\sigma}
							+4\s^2R_{\alpha'\sigma\beta'\sigma|\sigma} \right.\nonumber\\
							&\quad +g_{\alpha'\beta'}(\s^2-\r^2)R_{u\sigma u\sigma|\sigma}  - 4 \r \s^2\dot R_{\alpha'\sigma\beta'\sigma} \nonumber\\
							&\quad - 8\r \s^2R_{u(\alpha'\beta')\sigma|\sigma}  + 8\r^2\s^2\dot R_{u(\alpha'\beta')\sigma} \nonumber\\
							&\quad \left.+ 4\s^2 (\r^2+6\s^2)\left(R_{\alpha'u\beta'u|\sigma}-\r\dot R_{\alpha'u\beta'u}\right)\right]\bigg\} \nonumber\\
							&\quad +O(\lambda a^2,\lambda^2 a,\lambda^3).
\end{align}
\end{subequations}

We next move to the components of $h^{\S\S}_{tt}$, given in Eq.~\eqref{hSS_SC_Fermi}. In terms of tensors at $\bar x$, they are
\begin{subequations}
\begin{align}
h^{\S\S}_{tt} &= -\frac{m^2}{\lambda^2\bar\sigma}-\lambda^0\frac{7}{6\bar\sigma}R_{\bar u\bar\sigma\bar u\bar\sigma}
								+O(\lambda,a),\\
h^{\S\S}_{ta} &= \lambda^0m^2\frac{5e^\balpha_a}{3\bar\sigma}R_{\balpha\bsigma\bsigma\bu}+O(\lambda,a),\\
h^{\S\S}_{ab} &= \frac{m^2e^\balpha_a e^\bbeta_b}{4\bsigma^2}\bigg[\lambda^{-2}(10 g_{\balpha\bbeta} \bsigma 
								- 7 \sigma_{\balpha} \sigma_{\bbeta}) + \nonumber\\
							&\quad\lambda^0\Big(\tfrac{2}{15}g_{\balpha\bbeta}\bsigma R_{\bu\bsigma\bu\bsigma} 
								- \tfrac{16}{5} \bsigma R_{\balpha\bsigma\bbeta\bsigma} + \tfrac{104}{75} \bsigma^2R_{\balpha\bu\bbeta\bu} \nonumber\\
							&\quad + \tfrac{7}{5}\sigma_{\balpha} \sigma_{\bbeta} R_{\bu\bsigma\bu\bsigma}\Big)	
								-\frac{64}{15}\bsigma^2\ln(\lambda\sqrt{2\bar\sigma})R_{\bar u\balpha\bar u\bbeta}\bigg]\nonumber\\
							&\quad +O(\lambda\ln\lambda,a),
\end{align}
\end{subequations}
where we have again used Eq.~\eqref{ENN}. After an expansion about $x'$, the components become
\begin{subequations}\label{hSS_expansion}%
\allowdisplaybreaks\begin{align}
h^{\S\S}_{tt} &= -  \frac{2 m^2}{\lambda^2 \s^2} - \frac{m^2\lambda^0}{3\s^4}\left(2\r^2+7\s^2\right)R_{u\sigma u\sigma}+O(\lambda,a),\\
h^{\S\S}_{ta} &= - \frac{10 m^2\lambda^0e_a^{\alpha'}}{3\s^2}\left(R_{\alpha'\sigma u\sigma}+\r R_{\alpha'u\sigma u}\right)+O(\lambda,a),\\
h^{\S\S}_{ab} &= m^2e^{\alpha'}_ae^{\beta'}_b\bigg\{\frac{1}{\lambda^2\s^4}(5\s^2g_{\alpha'\beta'} 
								-7\sigma_{\alpha'} \sigma_{\beta'})\nonumber\\
								&\quad +\frac{\lambda^0}{\s^4}\bigl[ g_{\alpha'\beta'}\left(\tfrac{5}{3}\r^2+\tfrac{1}{15}\s^2\right) R_{u\sigma u\sigma}
								+\tfrac{26}{75}\s^4 R_{\alpha'u\beta'u} \nonumber\\
								&\quad + \tfrac{16}{5}\r\s^2 R_{u(\alpha'\beta')\sigma} 
								- \tfrac{8}{5}\r^2\s^2R_{\alpha'u\beta'u}  - \tfrac{8}{5} \s^2R_{\alpha'\sigma\beta'\sigma} \nonumber\\
								&\quad
								+ \tfrac{14}{3}\r\sigma_{(\alpha'}R_{\beta')\sigma u\sigma}  
								- \tfrac{7}{3}\r^2 \sigma_{(\alpha'}R_{\beta')u\sigma u} \nonumber\\
								&\quad + \left(\tfrac{7}{5}-\tfrac{14\r^2}{3\s^2}\right)R_{u\sigma u\sigma} \sigma_{\alpha'} \sigma_{\beta'}\bigr]
								 -\tfrac{16}{15}\ln(\lambda\s)R_{\alpha'u\beta'u}\bigg\}\nonumber\\
								&\quad +O(\lambda\ln\lambda,a).
\end{align}
\end{subequations}

Similarly, the components of $h^{\S\R}_{\mu\nu}$, given in Eq.~\eqref{hSR_SC_Fermi}, written in terms of tensors at $\bar x$ read
\begin{subequations}\label{hSR_barred}%
\begin{align}
h^{\S\R}_{tt} &= -\frac{mh^{\R1}_{\balpha\bbeta}}{\sqrt{2\bar\sigma}}
							\left(\frac{\sigma^\balpha\sigma^\bbeta}{2\bar\sigma}-\frac{1}{3}P^{\balpha\bbeta}\right)+O(\lambda^0),\\
h^{\S\R}_{ta} &= -\frac{mh^{\R1}_{\bgamma\bbeta}u^\bgamma e_a^\balpha}{\sqrt{2\bar\sigma}}
							\left(\frac{\sigma_\balpha\sigma^\bbeta}{2\bar\sigma}-\frac{1}{3}P_{\balpha}{}^{\bbeta}\right)+O(\lambda^0),\\
h^{\S\R}_{ab} &= \frac{me_a^\balpha e_b^\bbeta}{\sqrt{2\bar\sigma}}\Bigg[ 
							\frac{2h^{\R1}_{\bgamma(\balpha}\sigma_{\bbeta)}\sigma^{\bgamma}}{2\bar\sigma}
							-\frac{2}{3}h^{\R1}_{\bgamma(\balpha}P_{\bbeta)}{}^\bgamma\nonumber\\
							&\quad +P_{\balpha\bbeta}h^{\R1}_{\bar\mu\bar\nu}\left(\frac{2}{3}P^{\bar\mu\bar\nu}+\frac{1}{3}u^{\bar\mu}u^{\bar\nu}
								-\frac{\sigma^{\bar\mu}\sigma^{\bar\nu}}{2\bar\sigma}\right)\nonumber\\
							&\quad -\frac{\sigma_\balpha\sigma_\bbeta}{2\bar\sigma}h^{\R1}_{\bar\mu\bar\nu}
								\left(P^{\bar\mu\bar\nu}+u^{\bar\mu}u^{\bar\nu}\right)\Bigg]+O(\lambda^0),
\end{align}
\end{subequations}
and after an expansion around $x'$ they become
\begin{subequations}\label{hSR_expansion}%
\begin{align}
h^{\S\R}_{tt} &= -\frac{m}{\s^3}\left[h^{\R1}_{\sigma\sigma}+2\r h^{\R1}_{u\sigma}+\r^2h^{\R1}_{uu}
								-\frac{1}{3}\s^2 \left(h^{\R1}+h^{\R1}_{uu}\right)\right]\nonumber\\
							&\quad+O(\lambda^0),\\
h^{\S\R}_{ta} &= -\frac{me_a^{\alpha'}}{\s^3}\left[\left(h^{\R1}_{u\sigma}+\r h^{\R1}_{uu}\right)\sigma_{\alpha'}
								-\frac{1}{3}\s^2h^{\R1}_{u\alpha'}\right]+O(\lambda^0),\\
h^{\S\R}_{ab} &= \frac{me_a^{\alpha'} e_b^{\beta'}}{\s^3}\Bigg[ 
							2h^{\R1}_{\sigma(\alpha'}\sigma_{\beta')}+2\r h^{\R1}_{u(\alpha'}\sigma_{\beta')}
							-\frac{2}{3}\s^2h^{\R1}_{\alpha'\beta'}\nonumber\\
							&\quad +g_{\alpha'\beta'}\left(\frac{2}{3}\s^2h^{\R1}+\s^2 h^{\R1}_{uu}-h^{\R1}_{\sigma\sigma}
							-2\r h^{\R1}_{u\sigma}-\r^2h^{\R1}_{uu}\right)\nonumber\\
							&\quad -\sigma_{\alpha'}\sigma_{\beta'}\left(h^{\R1}+2h^{\R1}_{uu}\right)\Bigg]+O(\lambda^0),
\end{align}
\end{subequations}
where $h^{\R1}\equiv g^{\mu'\nu'}h^{\R1}_{\mu'\nu'}$. At this order in $\lambda$, $h^{\S\R}_{\mu\nu}$ depends on $h^{\R1}_{\mu\nu}$ only through its value on $\gamma$. However, at order $\lambda$, $h^{\S\R}_{\mu\nu}$ depends on partial derivatives of $h^{\R1}_{\mu\nu}$ on $\gamma$, and at order $\lambda^2$, it depends on second partial derivatives of $h^{\R1}_{\mu\nu}$ on $\gamma$. We can relate these derivatives to covariant quantities using the Christoffel symbols associated with the background metric~\eqref{background} in Fermi-Walker coordinates. The resulting relations are
\begin{subequations}\label{dhR}%
\begin{align}
\partial_a h^{\R1}_{\alpha\beta}(t,0) &= h^{\R1}_{\balpha\bbeta;\bgamma}e^\bgamma_a + O(a^\mu),\\
\partial_t h^{\R1}_{\alpha\beta}(t,0) &= h^{\R1}_{\balpha\bbeta;\bgamma}u^\bgamma + O(a^\mu),
\end{align}
\end{subequations}
and
\begin{subequations}\label{ddhR}%
\begin{align}
\partial_a\partial_b h^{\R1}_{\alpha\beta}(t,0) &= h^{\R1}_{\balpha\bbeta;\bgamma\bdelta}e^\bgamma_be^\bdelta_a
					+2R^{\bmu}{}_{b0a}u_{(\balpha}h^{\R1}_{\bbeta)\bmu}\nonumber\\
					&\quad -\frac{4}{3}R^{\bmu}{}_{(b\bnu)a}P^\bnu_{(\balpha}h^{\R1}_{\bbeta)\bmu} + O(a^\mu),\\
\partial_t\partial_a h^{\R1}_{\alpha\beta}(t,0) &= h^{\R1}_{\balpha\bbeta;\bgamma\bdelta}e^\bgamma_a u^\bdelta + O(a^\mu),\\
\partial_t\partial_t h^{\R1}_{\alpha\beta}(t,0) &= h^{\R1}_{\balpha\bbeta;\bgamma\bdelta}u^\bgamma u^\bdelta + O(a^\mu).
\end{align}
\end{subequations}

Last, we turn to $h^{\delta m}_{\mu\nu}$. Expanding Eq.~\eqref{dm_SC_Fermi} around $\tau'$ and simplifying yield
\begin{subequations}\label{dm_SC_expansion}%
\begin{align}
\delta m_{tt}(\bar\tau) &= - \tfrac{1}{3} m h^{\R1} - \tfrac{7}{3} m h_{uu}^{\R1}+O(\lambda),\\
\delta m_{ta}(\bar\tau) &= - \tfrac{4}{3} m e_a^{\alpha'}h_{u\alpha'}^{\R1}+O(\lambda), \\
\delta m_{ab}(\bar\tau) &= \tfrac{1}{3}m e_a^{\alpha'}e_b^{\beta'}\left[2h^{\R1}_{\alpha'\beta'} 
														+ g_{\alpha'\beta'}\left(h^{\R1}+3h^{\R1}_{uu}\right)\right]\nonumber\\
												&\quad+O(\lambda),
\end{align}
\end{subequations}
and
\begin{subequations}\label{hdm_SC_expansion}%
\begin{align}
h^{\delta m}_{tt} &= \frac{\delta m_{tt}(\tau')}{\s}+O(\lambda^0),\\
h^{\delta m}_{ta} &= \frac{\delta m_{ta}(\tau')}{\s}+O(\lambda^0),\\
h^{\delta m}_{ab} &= \frac{\delta m_{ab}(\tau')}{\s}+O(\lambda^0).
\end{align}
\end{subequations}

\subsubsection{Results: first order}\label{h1_SC_results}
In the self-consistent expansion, the first-order singular field in covariant form is obtained by combining Eqs.~\eqref{dt_expansion}, \eqref{dx_expansion}, \eqref{hS1_expansion}, and \eqref{ea.ea}. We write the result as a sum of two parts: acceleration-independent terms, and acceleration-dependent terms; that is,
\begin{equation}\label{hS1_SC_cov}
h^{\S1}_{\mu\nu} = h^{\S1\not a}_{\mu\nu}+h^{\S1 \bf{a}}_{\mu\nu}.
\end{equation}
The acceleration-independent terms are
\begin{align}\label{hS1_noa_cov}
h^{\S1\not a}_{\mu\nu} &=\frac{2 m}{\lambda\s} g^{\alpha'}_{\mu} g^{\beta'}_{\nu} \left(g_{\alpha'\beta'} + 2 u_{\alpha'} u_{\beta'}\right)\nonumber\\
&\quad + \lambda\frac{mg^{\alpha'}_\mu g^{\beta'}_{\nu}}{3\s^3} \Big[\!\left(\r^2 - \s^2\right)
			\left(g_{\alpha'\beta'}+2u_{\alpha'} u_{\beta'}\right)R_{u\sigma u\sigma} \nonumber\\
&\quad - 12\s^4 R_{\alpha' u\beta' u}- 12\r\s^2 u_{(\alpha'}R_{\beta')u\sigma u}\Big] \nonumber\\
&\quad + \lambda^2 \frac{mg^{\alpha'}_\mu g^{\beta'}_{\nu}}{12\s^3} \biggl\{16\r \s^2 u_{(\alpha'}R_{\beta')u\sigma u|\sigma} \nonumber\\
&\quad - 16\s^2\left(\r^2 + \s^2\right) u_{(\alpha'}\dot R_{\beta')u\sigma u} +(g_{\alpha'\beta'}+2 u_{\alpha'} u_{\beta'}) \nonumber\\
&\quad \times\left[\r (\r^2 - 3 \s^2)\dot R_{u\sigma u\sigma} + (\s^2- \r^2)R_{u\sigma u\sigma|\sigma}\right] \nonumber\\
&\quad + 24\s^4\left(R_{\alpha' u\beta'u|\sigma} - \r\dot R_{\alpha'u\beta'u}\right)  \biggr\}+O(\lambda^3),
\end{align}
in agreement with the results of, e.g., Heffernan~\emph{et al.}~\cite{Heffernan-etal:12} for the covariant expansion of the Detweiler-Whiting singular field~\cite{Detweiler-Whiting:03}; this demonstrates that at least through order $\lambda^2$, our singular-regular split of $h^1_{\mu\nu}$ agrees with the conventional split defined by Detweiler and Whiting. 

The acceleration-dependent terms are
\begin{align}\label{hS1_a_cov}
h^{\S1 \bf{a}}_{\mu\nu} &= \frac{m\lambda^0}{\s^3} g^{\alpha'}_\mu g^{\beta'}_{\nu} \left[
														\left(\s^2- \r^2\right) a_{\sigma} (g_{\alpha'\beta'}+2u_{\alpha'} u_{\beta'})\right.\nonumber\\
												&\quad \left.+8\r\s^2 a_{(\alpha'} u_{\beta')}\right] + \frac{m\lambda}{3\s^3}g^{\alpha'}_\alpha g^{\beta'}_{\beta} 
														\left\{12\s^2 (\r^2 + \s^2)\dot{a}_{(\alpha'}u_{\beta')}\right.\nonumber\\
												&\quad \left.
													 +  \r (3\s^2-\r^2)\dot{a}_{\sigma}(g_{\alpha'\beta'}+ 2u_{\alpha'} u_{\beta'})\right\}\nonumber\\
												&\quad	 +O(\lambda a^2,\lambda^2 a,\lambda^3).
\end{align}

\subsubsection{Results: second order}
The second-order singular field in the self-consistent expansion is the sum of three parts:
\begin{equation}\label{hS2_SC_cov}
h^{\S2}_{\mu\nu}=h^{\S\S}_{\mu\nu}+h^{\S\R}_{\mu\nu}+h^{\delta m}_{\mu\nu}.
\end{equation}

Combining Eqs.~\eqref{dt_expansion}, \eqref{dx_expansion}, and \eqref{ea.ea} with Eq.~\eqref{hSS_expansion} in~\eqref{hS_tensor}, we find
\begin{widetext}
\begin{align}\label{hSS_final}
h^{\S\S}_{\mu\nu}&= \lambda^{-2}\frac{m^2}{\s^4} g^{\alpha'}_{\mu} g^{\beta'}_{\nu} \Bigl\{ 5\s^2g_{\alpha'\beta'}
				-7 \sigma_{\alpha'}\sigma_{\beta'} - 14 \r \sigma_{(\alpha'} u_{\beta')}- (7 \r^2 - 3 \s^2) u_{\alpha'} u_{\beta'}\bigr]\Bigr\}
				- \frac{16}{15} m^2 g^{\alpha'}_{\mu} g^{\beta'}_{\nu} \ln(\lambda\s) R_{\alpha' u\beta' u} \nonumber\\
&\quad + \lambda^0\frac{m^2}{150 \s^6} g^{\alpha'}_{\mu} g^{\beta'}_{\nu} 
				\Bigg\{10\s^2 g_{\alpha'\beta'}\left(25 \r^2+\s^2\right)R_{\sigma u\sigma u}
			+ 20 \r \s^2 \left[35 \r \sigma_{(\alpha'} R_{\beta')u \sigma u} 
			+ \left(35 \r^2 - 31 \s^2\right) u_{(\alpha'} R_{\beta')u\sigma u} 
			- \s^2 R_{\sigma(\alpha' \beta') u}\right] \nonumber\\
&\quad + 10 \s^4 R_{\alpha' \sigma \beta' \sigma} - 350 \r\s^2\sigma_{(\alpha'}R_{\beta')\sigma u\sigma} 
			- 10\s^2\bigl(35 \r^2 - 17 \s^2\bigr) u_{(\alpha'}R_{\beta')\sigma u\sigma} 
			+ 2\s^4 \left(5 \r^2 + 26 \s^2\right)R_{\alpha' u\beta' u}\nonumber\\
&\quad - 70 \left[\left(10 \r^2 - 3 \s^2\right) \sigma_{\alpha'}\sigma_{\beta'}  
			+ 4 \r \left(5 \r^2 - 4 \s^2\right) u_{(\alpha'}\sigma_{\beta')}\right] R_{\sigma u\sigma u}  
			- 20\left(35 \r^4 - 53 \r^2 \s^2 - 6 \s^4\right) u_{\alpha'}u_{\beta'} R_{\sigma u\sigma u} \Bigg\}\nonumber\\
&\quad	+O\left(\lambda\ln \lambda,a^\mu\right).
\end{align}
Combining Eqs.~\eqref{dt_expansion}, \eqref{dx_expansion}, and \eqref{ea.ea} with Eq.~\eqref{hSR_expansion} in~\eqref{hS_tensor}, we find
\begin{align}\label{hSR_final}
h^{\S\R}_{\mu\nu} &= \lambda^{-1}\frac{m}{\s^3} g^{\alpha'}_{\mu} g^{\beta'}_{\nu} \Biggl\{g_{\alpha'\beta'}\left[\frac{2}{3}\s^2h^{\R1} 
			- \left(\r^2 -  \s^2\right)h^{\R1}_{uu} - h^{\R1}_{\sigma\sigma} - 2 \r h^{\R1}_{u\sigma}\right]-\frac{2}{3} h^{\R1}_{\alpha' \beta'} \s^2 
			+ 2h^{\R1}_{\sigma(\alpha'} \sigma_{\beta')} + 2\r h^{\R1}_{\sigma(\alpha'}u_{\beta')} \nonumber\\
&\quad - 2 h^{\R1}_{\sigma\sigma} u_{\alpha'} u_{\beta'}  -  h^{\R1} \Bigl[\sigma_{\alpha'}\sigma_{\beta'} + 2\r \sigma_{(\alpha'}u_{\beta')} 
			+ (\r^2 - \s^2) u_{\alpha'}u_{\beta'}\Bigr] + 2\r h^{\R1}_{u(\alpha'}\sigma_{\beta')} 
			+ 2(\r^2-\s^2)h^{\R1}_{u(\alpha'}u_{\beta')} \nonumber\\
&\quad	+ 4 h^{\R1}_{u\sigma} \sigma_{(\alpha'} u_{\beta')}- 2 h^{\R1}_{uu} \sigma_{\alpha'} \sigma_{\beta'}\Biggr\}  + O\left(\lambda^0\right).
\end{align}
\end{widetext}
Combining Eqs.~\eqref{dt_expansion}, \eqref{dx_expansion}, and \eqref{ea.ea} with Eqs.~\eqref{hdm_SC_expansion} and \eqref{dm_SC_expansion} in Eq.~\eqref{hS_tensor}, we find
\begin{equation}\label{hdm_SC_final}
h^{\delta m}_{\mu\nu} = \lambda^{-1}\frac{g^{\alpha'}_{\mu} g^{\beta'}_{\nu}\delta m_{\alpha' \beta'}}{\s} + O\left(\lambda^0\right),
\end{equation}
where
\begin{align}\label{dm_SC_cov}
\delta m_{\alpha\beta} &= \frac{1}{3}m\left(2h^{\R1}_{\alpha\beta}+g_{\alpha\beta}h^{\R1}\right)
				+4mu_{(\alpha}h^{\R1}_{\beta)\mu}u^\mu\nonumber\\
&\quad +m(g_{\alpha\beta}+2u_{\alpha} u_{\beta})u^\mu u^\nu h^{\R1}_{\mu\nu}.
\end{align}
In all of the above equations, $h^{\R1}_{\mu\nu}$ is evaluated on the accelerated worldline $\gamma$, and it is a functional of that worldline; it is not a functional of, for example, a geodesic tangential to $\gamma$.

The self-consistent equation of motion \eqref{motion_SC_Fermi} can be written in covariant form using \eqref{dhR}, leading to the standard expression 
\begin{equation}
a^\mu=\e F_1^\mu[\gamma] +O(\e^2),
\end{equation}
where
\begin{equation}\label{F1_SC}
F_1^\mu[\gamma]\equiv-\frac{1}{2}P^{\mu\nu}\left(2h^{\R1}_{\nu\lambda;\rho}[\gamma]-h^{\R1}_{\lambda\rho;\nu}[\gamma]\right)u^\lambda u^\rho
\end{equation}
is the self-force per unit mass.

Equations~\eqref{hSS_final}--\eqref{hdm_SC_final} are our main results for the covariant expansion of the second-order singular field in a self-consistent expansion. %They are of sufficiently high order to implement a self-consistent puncture scheme to calculate $h^{\R2}_{\mu\nu}$ on $\gamma$, evolving $\gamma$ using the equation of motion $a^\mu=F^\mu_1[\gamma]$, so long as the field is decomposed into $m$-modes (or tensor harmonics). 
Online, we make available terms through order $\lambda$ in all three of these equations~\cite{results}. For the quantities $h^{\S\S}_{\mu\nu}$ and $h^{\delta m}_{\mu\nu}$, we have analytically checked the results through order $\lambda$ by verifying that the expansions satisfy the correct wave equations through order $1/\lambda$. For the quantity $h^{\S\R}_{\mu\nu}$, we have analytically checked the results through order $\lambda^0$ by verifying that the expansions satisfy the correct wave equations through order $1/\lambda^2$; we have not verified the order-$\lambda$ term in $h^{\S\R}_{\mu\nu}$ in this way, due simply to the algebraic complexity involved. As we discuss in Sec.~\ref{required_order} below, in most practical schemes this term will not be needed anyway.

\subsection{Gralla-Wald form}
We now turn to the covariant expansion in the Gralla-Wald case. As in Fermi-Walker coordinates, nearly all the results can be recycled from the self-consistent expansion simply by setting $a^\mu=0$ and making the replacement $\gamma\to\gamma_0$ in the functional dependence of $h^{\R1}_{\mu\nu}$. The points $\bar x$ and $x'$ in this case refer to points $z_0^\mu(\bar\tau)$ and $z_0^\mu(\tau')$ on $\gamma_0$.

\subsubsection{Expansions of $h^\S_{tt}$, $h^\S_{ta}$, and $h^\S_{ab}$}
Our result~\eqref{hS1_expansion} for the components of the first-order field can be used directly, simply by setting $a^\mu=0$. Our results~\eqref{hSS_expansion} and \eqref{hSR_expansion} for the components of $h^{\S\S}_{\mu\nu}$, in which we already neglected acceleration terms, can be used after making the single change $h^{\R1}_{\mu\nu}[\gamma]\to h^{\R1}_{\mu\nu}[\gamma_0]$. The formula~\eqref{hdm_SC_expansion} for the components of $h^{\delta m}_{\mu\nu}$ remains intact, but the components of $\delta m_{\mu\nu}$ on the right-hand side are modified by the inclusion of the $z_1^a$ term in Eq.~\eqref{dm_GW_Fermi}:
\begin{subequations}\label{dm_GW_expansion}%
\begin{align}
\delta m_{tt}(\bar\tau) &= - \tfrac{1}{3} m h^{\R1} - \tfrac{7}{3} m h_{uu}^{\R1}+O(\lambda),\\
\delta m_{ta}(\bar\tau) &= - 4me_a^{\alpha'}\left(\tfrac{1}{3} h_{u\alpha'}^{\R1}+\dot z_{1\alpha'}^\perp\right)+O(\lambda), \\
\delta m_{ab}(\bar\tau) &= \tfrac{1}{3}m e_a^{\alpha'}e_b^{\beta'}\left[2h^{\R1}_{\alpha'\beta'} 
														+ g_{\alpha'\beta'}\left(h^{\R1}+3h^{\R1}_{uu}\right)\right]\nonumber\\
												&\quad+O(\lambda),
\end{align}
\end{subequations}
where $\dot z_{1\alpha}^\perp\equiv \frac{D\dot z_{1\alpha}^\perp}{d\tau}$.

This leaves the components of $h^{\delta z}_{\mu\nu}$. We rewrite Eq.~\eqref{hdz_Fermi} in terms of tensors at $\bar x$, yielding
\begin{align}
h^{\delta z}_{tt} &= -\frac{2mz^\perp_{1\bar\alpha}\sigma^{\bar\alpha}}{(2\bar\sigma)^{3/2}}+O(\lambda^0),\\
h^{\delta z}_{ta} &= O(\lambda^0),\\
h^{\delta z}_{ab} &= -\frac{2mz^\perp_{1\bar\alpha}\sigma^{\bar\alpha}}{(2\bar\sigma)^{3/2}}+O(\lambda^0),
\end{align}
and we then expand around $x'$, yielding
\begin{align}\label{hdz_expansion}
h^{\delta z}_{tt} &= -\frac{2m(z^\perp_{1\alpha'}+\lambda\r\dot z^\perp_{1\alpha'})\sigma^{\alpha'}}{\s^3}+O(\lambda^0),\\
h^{\delta z}_{ta} &= O(\lambda^0),\\
h^{\delta z}_{ab} &= -\frac{2m(z^\perp_{1\alpha'}+\lambda\r\dot z^\perp_{1\alpha'})\sigma^{\alpha'}}{\s^3}+O(\lambda^0).
\end{align}

\subsubsection{Results: first order}
The first-order singular field in the Gralla-Wald-type expansion is given by 
\begin{equation}\label{hS1_GW_cov}
h^{\S1}_{\mu\nu}=h^{\S1\not a}_{\mu\nu},
\end{equation}
where $h^{\S1\not a}_{\mu\nu}$ is as it appears in Eq.~\eqref{hS1_noa_cov}, with $x'$ now referring to a point on $\gamma_0$.

\subsubsection{Results: second order}
We write the second-order singular field in the Gralla-Wald-type expansion as the sum of four parts:
\begin{equation}\label{hS2_GW_cov}
h^{\S2}_{\mu\nu} = h^{\S\S}_{\mu\nu}+h^{\S\R}_{\mu\nu}+h^{\delta m}_{\mu\nu}+h^{\delta z}_{\mu\nu}.
\end{equation}
$h^{\S\S}_{\mu\nu}$, $h^{\S\R}_{\mu\nu}$, and $h^{\delta m}_{\mu\nu}$ are given by Eqs.~\eqref{hSS_final}, \eqref{hSR_final}, and \eqref{hdm_SC_final}, with $x'$ now a point on $\gamma_0$, $u^{\alpha'}$ replaced by $u_0^{\alpha'}$, $a^\mu$ set to zero, $h^{\R1}_{\mu\nu}[\gamma]$ replaced by $h^{\R1}_{\mu\nu}[\gamma_0]$, and Eq.~\eqref{dm_SC_cov} replaced by
\begin{align}\label{dm_GW_cov}
\frac{\delta m_{\alpha\beta}}{m} &= \frac{1}{3}\left(2h^{\R1}_{\alpha\beta}+g_{\alpha\beta}h^{\R1}\right)\nonumber\\
&\quad +(g_{\alpha\beta}+2u_{0\alpha} u_{0\beta})u_0^\mu u_0^\nu h^{\R1}_{\mu\nu}\nonumber\\
&\quad +4u_{0(\alpha}\left(h^{\R1}_{\beta)\mu}u_0^\mu+2\dot z^\perp_{1\beta)}\right).
\end{align}

The new term (as compared to the self-consistent expansion) is $h^{\delta z}_{\mu\nu}$. Combining Eqs.~\eqref{dt_expansion}, \eqref{dx_expansion}, and \eqref{ea.ea} with Eq.~\eqref{hdz_expansion} via~\eqref{hS_tensor}, we find
\begin{align}\label{hdz_final}
h^{\delta z}_{\mu\nu}&= -\frac{2mg^{\alpha'}_{\mu} g^{\beta'}_{\nu} \left(g_{\alpha'\beta'} + 2 u_{0\alpha'}u_{0\beta'}\right)}{\lambda^2\s^3}
	\left(z_{1\perp}^{\gamma'}+\lambda \r \dot z_{1\perp}^{\gamma'}\right)\sigma_{\gamma'} \nonumber\\
&\quad + O\left(s^0\right).
\end{align}
The evolution of $z^\alpha_{1\perp}$ is governed by the Gralla-Wald equation~\eqref{motion_GW_Fermi}, which we rewrite in covariant form as
\begin{equation}
\frac{D^2 z^{\alpha}_{1\perp}}{d\tau^2} = -R^\alpha{}_{\mu\beta\nu}u^\mu_0 z^{\beta}_{1\perp} u_0^\nu + F_1^\alpha[\gamma_0],\label{motion}
\end{equation}
where
\begin{equation}\label{F1_GW}
F^\mu_1[\gamma_0]=-\frac{1}{2}P_0^{\mu\nu}\left(2h^{\R1}_{\nu\lambda;\rho}[\gamma_0]-h^{\R1}_{\lambda\rho;\nu}[\gamma_0]\right)u_0^\lambda u_0^\rho.
\end{equation}

Equation~\eqref{hS2_GW_cov}, with the first three terms as already given in the self-consistent case, and Eq.~\eqref{hdz_final} are our main results for the covariant expansion of the second-order singular field in our Gralla-Wald-type expansion. Online, we make available terms through order $\lambda$ in Eq.~\eqref{hdz_final}~\cite{results}, and we have analytically checked the result through that order by verifying that it satisfies the correct wave equation through order $1/\lambda$.

%%%%%%%%%%%%%%%%%%%%%%%%%%%%%%%%%%%%%%%%%%%%%%%%%%%%%
\section{Conclusion: steps toward practical implementation}
%%%%%%%%%%%%%%%%%%%%%%%%%%%%%%%%%%%%%%%%%%%%%%%%%%%%%
We have derived covariant expansions of the second-order singular field in an arbitrary vacuum background, in both a self-consistent formalism and a Gralla-Wald-type formalism. Our final results in the self-consistent case are Eqs.~\eqref{hS1_SC_cov}--\eqref{hdm_SC_final}; in the Gralla-Wald case, they are Eqs.~\eqref{hS1_GW_cov}--\eqref{F1_GW}. We have also made higher-order terms in the expansions available online~\cite{results}.

To make use of these results in practice, a few steps must be taken.

\subsection{Puncture as a coordinate expansion}
For a practical numerical implementation of a puncture scheme, the puncture must be written in a specified coordinate system. This might mean that the expansion of the singular field must be written directly in the coordinates one will use in one's numerical evolution. Or if one wishes to decompose the puncture into a useful basis of functions, such as tensor harmonics in Schwarzschild, it might mean that the expansion should be written in some coordinate system convenient for the calculation of that decomposition, as was done in the frequency-domain puncture scheme of Warburton and Wardell~\cite{Warburton-Wardell:14}.

Expressing the expansion in coordinate form is, in principle, straightforward. The covariant expansion of the singular field can be recast as an expansion in coordinate differences $\Delta x^{\alpha'}=x^\alpha-x^{\alpha'}$, where $x^{\alpha'}$ are the coordinate values at the point $x'$ on the worldline  (by which we mean $\gamma$ in the self-consistent case, $\gamma_0$ in the Gralla-Wald case). All that is required is the expansion of the covariant quantities $\sigma^{\alpha'}$ and $g^{\alpha'}_\beta$ in powers of $\Delta x^{\alpha'}$. Following Ref.~\cite{Heffernan-etal:12}, the expansion of $\sigma_{\alpha'}$ can be found by writing
\begin{align}
\sigma(x,x') &= \frac{1}{2}g_{\alpha'\beta'}\Delta x^{\alpha'}\Delta x^{\beta'} 
			+ A_{\alpha'\beta'\gamma'}\Delta x^{\alpha'}\Delta x^{\beta'}\Delta x^{\gamma'} \nonumber\\
			&\quad + B_{\alpha'\beta'\gamma'\delta'}\Delta x^{\alpha'}\Delta x^{\beta'}\Delta x^{\gamma'}\Delta x^{\delta'} +\ldots, %\nonumber\\
%			&\quad + C_{\alpha'\beta'\gamma'\delta'\zeta'}\Delta x^{\alpha'}\Delta x^{\beta'}\Delta x^{\gamma'}
%					\Delta x^{\delta'}\Delta x^{\zeta'}\nonumber\\
%			&\quad +O[(\Delta x^{\alpha'})^6],\\
\end{align}
then acting with a partial derivative on the equation, and finally determining the coefficients in the expansions recursively by using the identity $\sigma^{\alpha'}\sigma_{\alpha'}=2\sigma(x,x')$. Similarly, the expansion of $g^{\alpha'}_{\beta}$ can be found by writing
\begin{align}
g^{\alpha'}_\beta &= \delta^{\alpha'}_{\beta'}+G^{\alpha'}{}_{\beta'\gamma'}\Delta x^{\gamma'}
							+G^{\alpha'}{}_{\beta'\gamma'\delta'}\Delta x^{\gamma'}\Delta x^{\delta'}+\ldots,
\end{align}
%\sigma_{\alpha'} &= g_{\alpha'\beta'}\Delta x^{\beta'} 
%			+ (g_{\beta'\gamma',\alpha'}+3A_{\alpha'\beta'\gamma'})\Delta x^{\beta'}\Delta x^{\gamma'} \nonumber\\
%			&\quad + (A_{\beta'\gamma'\delta',\alpha'}+4B_{\alpha'\beta'\gamma'\delta'})\Delta x^{\beta'}
%			\Delta x^{\gamma'}\Delta x^{\delta'}\nonumber\\
%			&\quad +(B_{\beta'\gamma'\delta'\zeta',\alpha'}+5C_{\alpha'\beta'\gamma\delta'\zeta'})
%			\Delta x^{\beta'}\Delta x^{\gamma'}\Delta x^{\delta'}\Delta x^{\zeta'}\nonumber\\
%			&\quad +O[(\Delta x^{\alpha'})^5] 
acting with a partial derivative, and then determining the coefficients using the identity $g^{\alpha'}_{\beta;\gamma'}\sigma^{\gamma'}=g^{\alpha'}_{\beta,\gamma'}\sigma^{\gamma'}+\Gamma^{\alpha'}_{\gamma'\delta'}g^{\delta'}_\beta\sigma^{\gamma'}=0$.

The end result will be an expansion of the form~\eqref{hP2_coords}~\cite{Pound:13b}. To aid the discussion in the following section, we rewrite that result more transparently as
\begin{subequations}\label{coord_expansions}%
\begin{align}
h^{\S\S}_{\mu\nu} & \sim \frac{(\Delta x)^2}{\lambda^{2}\rho^4}+\frac{(\Delta x)^5}{\lambda\rho^6}
								+\lambda^{0}\frac{(\Delta x)^8}{\rho^8}+\lambda\frac{(\Delta x)^{11}}{\rho^{10}}\nonumber\\
									&\quad +\left[\lambda^0(\Delta x)^0+\lambda(\Delta x)^1\right]\ln(\lambda\rho)\nonumber\\
									&\quad +O(\lambda^2\ln\lambda),\\
h^{\delta z}_{\mu\nu}& \sim \frac{(\Delta x)^1}{\lambda^{2}\rho^3}+\frac{(\Delta x)^4}{\lambda\rho^5}
								+\lambda^0\frac{(\Delta x)^{7}}{\rho^{7}}+\lambda\frac{(\Delta x)^{10}}{\rho^{9}}\nonumber\\
									&\quad+O(\lambda^2),\\
h^{\S\R}_{\mu\nu} & \sim \frac{(\Delta x)^2}{\lambda\rho^3}+\lambda^{0}\frac{(\Delta x)^5}{\rho^5}
								+\lambda\frac{(\Delta x)^{8}}{\rho^{7}}+O(\lambda^2),\\
h^{\delta m}_{\mu\nu}& \sim \frac{(\Delta x)^0}{\lambda\rho}+\lambda^{0}\frac{(\Delta x)^3}{\rho^3}
								+\lambda\frac{(\Delta x)^{6}}{\rho^{5}}+O(\lambda^2),
\end{align}
\end{subequations}
where `$(\Delta x)^n$' indicates a polynomial in $\Delta x^{\mu'}$ of homogeneous order $n$. Each polynomial is of the form $P_{\mu'\nu'\alpha'_1\cdots\alpha'_n}(x')\Delta x^{\alpha'_1}\cdots\Delta x^{\alpha'_n}$ with some coefficient $P_{\mu'\nu'\alpha'_1\cdots\alpha'_n}(x')$ that depends only on $x'$. One can easily derive the general structure of the expansion~\eqref{coord_expansions} by substituting generic power expansions $\sigma_{\mu'}\sim \sum_{n\geq0} \lambda^n(\Delta x)^n$ and $g^{\mu'}_\mu\sim\sum_{n\geq0} \lambda^n(\Delta x)^n$ into the covariant expansions of $h^{\S2}_{\mu\nu}$. We have simplified the results by obtaining a common denominator at each order in $\lambda$, using the fact that $\rho^2\sim (\Delta x)^2$.

\subsection{Required order of the puncture}\label{required_order}
Before implementation, one must also decide how many orders in distance should be included in the puncture for one's particular purposes. We cursorily described the requisite orders in Sec.~\ref{strategy}; we explain them more thoroughly here.

To calculate the second-order force, one requires $\partial h^{\res 2}_{\mu\nu}=\partial h^{\R 2}_{\mu\nu}$ on the worldline. This means we must have $\lim_{x\to\gamma}\left(\partial h^{\P 2}_{\mu\nu}-\partial h^{\S 2}_{\mu\nu}\right)=0$, or in other words, $\partial h^{\P 2}_{\mu\nu}-\partial h^{\S 2}_{\mu\nu}=o(\lambda^0)$. From this one might infer that for the purpose of calculating the second-order force, $h^{\P 2}_{\mu\nu}$ must include all terms in $h^{\S 2}_{\mu\nu}$ through order $\lambda$. If one were to implement a puncture scheme in 3+1D, that would be true. 

However, analysis has shown~\cite{Barack-Golbourn-Sago:07} that in a puncture scheme that decomposes the field into azimuthal $m$-modes $e^{im \phi}$, one can sometimes lower the required order of the puncture by one power. (Of course, the same statements also hold true if one performs a complete tensor-harmonic decomposition rather than an $m$-mode decomposition alone.) % The reason, roughly for this can be most easily understood in the context of a full spherical harmonic expansion of the angular dependence. At a point of discontinuity, a series $f=\sum f_{lm}Y^{lm}$ converges to the average of $f$ around an infinitesimal circle around the point~\cite{Sansone:77}. Therefore, if $f$ has odd parity about the point of discontinuity, the sum vanishes. 
Specifically, one can neglect an order-$\lambda^0$ term, even though it is finite in the limit to the worldline, if it has odd parity about the worldline. By odd parity we mean a change of sign under the parity transformation $\Delta x^{\mu'}\to-\Delta x^{\mu'}$. So, for example, a term like $\Delta x^{\mu'}/\sqrt{P_{\alpha'\beta'}\Delta x^{\alpha'}\Delta x^{\beta'}}$ can be dropped from one's puncture. The reason this is allowed can be understood intuitively from the fact that the $m$-mode decomposition of a function converges to the function's average across the point of discontinuity; therefore in the limit to the worldline, these odd-parity terms contribute nothing to the decomposed puncture. One can show that the regular field at a point on the worldline can then be calculated as the sum over modes of the residual field at that point. 

Unlike in the first-order case, where all terms in the singular field at a given order in $\lambda$ share the same parity, %In the first-order case, the order-$\lambda$ this means that so long as we decompose the field at least into $m$-modes, to calculate the  (of course, decomposition into tensor harmonics would work as well) all the order-$\lambda^0$ terms have odd parity, and they need not be included in the puncture. Similarly, if we wish to calculate the first-order force, we need not include any terms of order $\lambda$ in the puncture (since the parity is reversed at each successive order in $h^{\S1}_{\mu\nu}$ and derivatives lower the order of a given term by one and reverse its parity). 
in the second-order case different pieces of the field have different parities: the order-$\lambda^0$ terms
\begin{itemize}
\item in $h^{\S\S}_{\mu\nu}$ have even parity,
\item in $h^{\S\R}_{\mu\nu}$, $h^{\delta z}_{\mu\nu}$, and $h^{\delta m}_{\mu\nu}$ have odd parity,
\end{itemize}
and at successive orders in $\lambda$ the parity alternates. These properties are  made obvious in Eq.~\eqref{coord_expansions}. (They can also be inferred from the effect of the parity transformation $n^i\to-n^i$ on Eq.~\eqref{hS2_form_Fermi}, or from that of $\sigma^{\alpha'}\to-\sigma^{\alpha'}$ on Eq.~\eqref{hS2_form_cov}; the parity in all three cases will be the same~\cite{Pound-Merlin-Barack:14}.)

Therefore, assuming at least an $m$-mode decomposition, to calculate $h^{\R2}_{\mu\nu}$ on the worldline one must include in one's puncture the order-$\lambda^0$ terms from $h^{\S\S}_{\mu\nu}$, but one need not include any of the order-$\lambda^0$ terms from $h^{\S\R}_{\mu\nu}$, $h^{\delta z}_{\mu\nu}$, or $h^{\delta m}_{\mu\nu}$. Similarly, since differentiation both reduces the order of a term and reverses its parity, to calculate the second-order force one must include the order-$\lambda$ terms from $h^{\S\S}_{\mu\nu}$, but one need not include those from $h^{\S\R}_{\mu\nu}$, $h^{\delta z}_{\mu\nu}$, or $h^{\delta m}_{\mu\nu}$. 

In the body of this paper we have presented results for $h^{\S\S}_{\mu\nu}$ through order $\lambda^0$ and $h^{\S\R}_{\mu\nu}$, $h^{\delta z}_{\mu\nu}$, and $h^{\delta m}_{\mu\nu}$ through order $1/\lambda$. Due to the savings in a mode decomposition, these results are of sufficiently high order to calculate the second-order regular field on the worldline. In the self-consistent case, implementing a scheme using this puncture would consist of solving the wave equations~\eqref{h1_SC} and \eqref{h2_SC} with the puncture following a trajectory $\gamma$ governed by the first-order equation of motion
\begin{align}
\frac{D^2 z^\mu}{d\tau^2} &= -\frac{1}{2}P^{\mu\gamma}
		\left(2h^{\res1}_{\gamma\alpha;\beta}-h^{\res1}_{\alpha\beta;\gamma}\right)u^\alpha u^\beta,
\end{align}
and the second-order regular field would be calculated as $h^{\R2}_{\mu\nu}=h^{\res2}_{\mu\nu}$ on $\gamma$ (with $h^{\res2}_{\mu\nu}$ defined as the sum over its modes). In the Gralla-Wald case, the scheme would consist of solving the sequence of equations \eqref{h1_GW}, \eqref{motion_GW}, and \eqref{h2_GW}, and the second-order regular field would be calculated as $h^{\R2}_{\mu\nu}=h^{\res2}_{\mu\nu}$ on the reference geodesic $\gamma_0$.

The results we have made available online are of sufficient order to calculate the second-order force even in 3+1D, and we have stringently tested their correctness through at least the order required to do the same in an $m$-mode scheme: order $\lambda$ for $h^{\S\S}_{\mu\nu}$ and order $\lambda^0$ for $h^{\S\R}_{\mu\nu}$, $h^{\delta z}_{\mu\nu}$, and $h^{\delta m}_{\mu\nu}$. With a puncture of that order in the self-consistent case, the wave equations~\eqref{h1_SC} and \eqref{h2_SC} can be solved with the puncture moving according to the second-order equation of motion~\eqref{motion_SC}. This scheme should maintain second-order accuracy on a timescale $\sim1/\e$, whereas the scheme using the first-order equation of motion can be expected to be uniform only on the shorter timescale $\sim1/\sqrt{\e}$, based on the error estimate $\delta z\sim \delta a\, t^2$, where $\delta z$ is the error in position and $\delta a$ the error in acceleration. In the Gralla-Wald case, calculating the second-order force would allow one to calculate the second-order correction to the position, $z_2^\alpha$.

\subsection{Order-reduction of the self-consistent system}\label{order-reduction}
One major remaining point to consider pertains to the handling of acceleration terms in the self-consistent formalism. It is well known that self-consistent derivations of equations of motion often lead to ill-behaved third-order-in-time differential equations. The most famous example of this is the Abraham-Lorentz-Dirac equation for a charged particle. In the case of a small mass, if we write the first-order-accurate equation of motion in the form
\begin{equation}\label{a1-second-order-in-time}
a^\mu=-\frac{1}{2}P^{\mu\nu}\left(2h^{\R1}_{\nu\lambda;\rho}[\gamma]-h^{\R1}_{\lambda\rho;\nu}[\gamma]\right)u^\lambda u^\rho,
\end{equation}
and we write the field $h^1_{\mu\nu}[\gamma]$ in terms of a Green's function $G_{\mu\nu\mu'\nu'}$ as 
\begin{equation}
h^1_{\mu\nu} = 2m\int_\gamma G_{\mu\nu\mu'\nu'}(g^{\mu'\nu'}+2u^{\mu'}u^{\nu'})d\tau,
\end{equation}
and we then expand $h^1_{\mu\nu}$ near $\gamma$ and identify the contributions to $h^{\R1}_{\mu\nu}[\gamma]$, then we find
\begin{equation}\label{a1-third-order-in-time}
a^\mu = e^{a\mu}\left(-h^{\rm tail}_{0a0}+\tfrac{1}{2}h^{\rm tail}_{00a}-\tfrac{11}{3}m\dot a_a\right),
\end{equation}
where the tail terms at time $\tau$ are defined as
\begin{equation}
h^{\rm tail}_{0a0}\equiv u^\alpha e^\beta_a u^\gamma 2m \int_{-\infty}^{\tau-0^+} G_{\mu\nu\mu'\nu';\gamma}(g^{\mu'\nu'}+2u^{\mu'}u^{\nu'})d\tau
\end{equation}
and analogously for $h^{\rm tail}_{00a}$. These results can be easily derived from the explicit results for $h^{\R1}_{\mu\nu}(r=0)$ and $\partial_\rho h^{\R1}_{\mu\nu}(r=0)$ in Table I of Ref.~\cite{Pound:10b}.\footnote{However, note that Table I in Ref.~\cite{Pound:10b} is missing a factor of 4 from the $ma_a$ term in the quantity $\hat C_a^{(1,0)}=h^{\R1}_{ta}(r=0)$. The missing 4 appears in its correct location in Eq.~(E.9) of that reference.} 

The $\dot a_i$ term in Eq.~\eqref{a1-third-order-in-time} is the gravitational antidamping term discovered by Havas~\cite{Havas:57} (as corrected by Havas and Goldberg~\cite{Havas-Goldberg:62}). Its presence has made the apparently second-order-in-time differential equation~\eqref{a1-second-order-in-time} into an apparently third-order-in-time integro-differential equation. An important question is whether this feature manifests in the coupled system we would hope to solve numerically, made up of Eqs.~\eqref{h1_SC}--\eqref{motion_SC}. The answer would seem to be that the problem has been shifted elsewhere: the acceleration and its time derivatives now appear in the source term $E_{\mu\nu}[h^{\P}]$ in the field equation. If we imagine solving the coupled system at a given time step, we can see that we would need to know the acceleration at that time step in order to calculate the field, but we would need to know the value of that same field before we could calculate the acceleration. One possibility might be to solve this problem iteratively at each time step. But a much simpler alternative would be to effectively perform a reduction-of-order procedure on the wave equations. Noting that $a^\mu\sim\e$, we can see that we would still preserve second-order accuracy by moving the acceleration-dependent terms from the first-order puncture into the second-order one; indeed, we have already done an analogous thing by neglecting explicit acceleration terms in our second-order puncture. Furthermore, we can then see that replacing $a^\mu$ with $F_1^\mu[\gamma]$ would also preserve the desired accuracy. Therefore, we can shift the term $h^{\S1\bf a}_{\mu\nu}$ from Eq.~\eqref{hS1_SC_cov} into Eq.~\eqref{hS2_SC_cov}, such that
\begin{equation}
h^{\P1}_{\mu\nu} = h^{\S1\not a}_{\mu\nu}
\end{equation}
(with an implied truncation of the right-hand side at a specified order in $\lambda$) and
\begin{equation}
h^{\P2}_{\mu\nu} = h^{\S\S}_{\mu\nu}+h^{\S\R}_{\mu\nu}+h^{\delta m}_{\mu\nu}+h^{\S1\bf a}_{\mu\nu}.
\end{equation}
In $h^{\S1\bf a}_{\mu\nu}$ one can then make the replacement $a^\mu\to F_1^\mu[\gamma]$, with $F_1^\mu[\gamma]$ given by Eq.~\eqref{F1_SC}. This alteration offers a substantial simplification of the coupled field-motion system, with the acceleration and its derivatives appearing nowhere except on the left-hand side of Eq.~\eqref{motion_SC}.

\subsection{Prospectus}
In the future we expect both our self-consistent and Gralla-Wald-type results to be of use in practice, but for differing purposes. While the self-consistent scheme offers the prospect of long-term accuracy, it has the drawback of requiring an evolution in the time domain: since the trajectory sourcing the field responds dynamically to that field, there is no clear way to avoid solving the coupled field-motion equations time-step by time-step. Therefore, in order to take advantage of the long-term accuracy provided by the self-consistent approximation, one must also achieve the feat of maintaining numerical accuracy on those long timescales of $\sim 10^6$ orbits. For that reason, the Gralla-Wald-type scheme, despite its obvious drawback of being valid only on timescales much shorter than an inspiral, will be preferable for many purposes, such as calculating short-term conservative effects and fixing parameters in effective-one-body theory. Furthermore, a Gralla-Wald-type scheme has the distinct advantage of being amenable to treatment in the frequency domain, at least for certain calculations~\cite{Pound:14c}. Warburton and Wardell have recently devised a frequency-domain puncture scheme that should be generalizable to this case~\cite{Warburton-Wardell:14}.

However, in extreme cases, such as zoom-whirl orbits in Schwarzschild, which lie near the separatrix between bound and unbound orbits, a self-consistent calculation may be unavoidable, because in such cases the reference geodesic may diverge very rapidly from the accelerated orbit, and the correction $z_1^\mu$ to the position may grow exponentially quickly. The self-consistent formalism also allows one to more easily derive alternative, but more easily implemented, approximation schemes that preserve long-term accuracy; for example, by starting from the self-consistent equations, one can readily derive a two-timescale expansion of the coupled field-motion problem~\cite{Hinderer-Flanagan:08,Pound:14d}, which should be accurate over a complete inspiral without requiring long-term evolutions in the time-domain.

\begin{acknowledgments}
The calculation of a covariant puncture was first suggested to us by Eric Poisson. We also thank Leor Barack for helpful discussions and comments. The research leading to these results received funding from the European Research Council under the European Union's Seventh Framework Programme (FP7/2007-2013)/ERC grant agreement no. 304978. AP also acknowledges support from the Natural Sciences and Engineering Research Council of Canada.
\end{acknowledgments}

\appendix
\section{Relationship between two definitions of the regular field}\label{trace-reverse}
In the present paper, we adopted the definition of the regular field from Ref.~\cite{Pound:12a}, as described in Sec.~\ref{singular-regular}. In Ref.~\cite{Pound:12b}, one of us presented results based on the same style of definition, but applied to the trace-reversed second-order field $\bar h^{\mu\nu}_2$. These two definitions differ from one another. In this appendix we show the relationships between them at the first two orders of an expansion in Fermi-Walker coordinates. Let us refer to the definition used in this paper as definition A, and that used in Ref.~\cite{Pound:12b} as definition B.

We begin by decomposing the pieces of the regular field in definition A into irreducible STF pieces, following the notation in Refs.~\cite{Pound:10a,Poisson-Pound-Vega:11}:
\begin{align}
h^{\R 2}_{tt}(t,0) &\equiv h^{(2,0,0,0)}_{tt} = \hat A^{(2,0)},\\
h^{\R 2}_{ta}(t,0) &\equiv h^{(2,0,0,0)}_{ta} = \hat C^{(2,0)}_{a},\\
h^{\R 2}_{ab}(t,0) &\equiv h^{(2,0,0,0)}_{ab} = \delta_{ab}\hat K^{(2,0)} + \hat H^{(2,0)}_{ab},\\
h^{\R 2}_{tt,i}(t,0) &\equiv h^{(2,1,0,1)}_{tti} = \hat A^{(2,1)}_i,\\
h^{\R 2}_{ta,i}(t,0) &\equiv h^{(2,1,0,1)}_{tai} = \hat C^{(2,1)}_{ai} + \epsilon^b{}_{ai}\hat D^{(2,1)}_{b} + \delta_{ai}\hat B^{(2,1)},\\
h^{\R 2}_{ab,i}(t,0) &\equiv h^{(2,1,0,1)}_{abi} = \delta_{ab}\hat K^{(2,1)}_i + \hat H^{(2,1)}_{abi} +\epsilon_i{}^c{}_{(a}\hat I^{(2,1)}_{b)c} \nonumber\\
						&\hphantom{\equiv h^{(2,1,0,1)}_{abi} =} +\delta_{i\langle a}\hat F^{(2,1)}_{b\rangle},
\end{align}
where each of the hatted quantites is an STF Cartesian 3-tensor. At these orders, definition B states 
\begin{align}
\bar h_{\R 2}^{tt}(t,0) &\equiv \bar h_{(2,0,0,0)}^{tt},\\
\bar h_{\R 2}^{ta}(t,0) &\equiv \bar h_{(2,0,0,0)}^{ta},\\
\bar h_{\R 2}^{ab}(t,0) &\equiv \bar h_{(2,0,0,0)}^{ab},\\
\bar h_{\R 2}^{tt,i}(t,0) &\equiv \bar h_{(2,1,0,1)}^{tti},\\
\bar h_{\R 2}^{ta,i}(t,0) &\equiv \bar h_{(2,1,0,1)}^{tai},\\
\bar h_{\R 2}^{ab,i}(t,0) &\equiv \bar h_{(2,1,0,1)}^{abi}.
\end{align}
The latter expressions are not the trace reverse of the former, but rather independent definitions based on picking out particular coefficients in the two expansions $\sum r^p (\ln r)^qh^{(npq\ell)}_{\mu\nu L}\nhat^L$ and $\sum r^p (\ln r)^q\bar h_{(npq\ell)}^{\mu\nu L}\nhat_L$.

By taking the trace-reverse of $\sum r^p (\ln r)^q\bar h_{(npq\ell)}^{\mu\nu L}\nhat_L$ and decomposing the result into irreducible pieces, we find the following relations:
\begin{align}
\hat A^{(2,0)} &= -\frac{59}{6} m^2 a{}_{a} a{}^{a} + \frac{1}{2} \bar h^{a}_{\R2}{}_{a} + \frac{1}{2} \bar h^{tt}_{\R2},\\
\hat C^{(2,0)}_a &= - \bar h^{t}_{\R2}{}_{a},\\
\hat H^{(2,0)}_{ab} &= -\frac{5}{9} m^2 \mathcal{E}{}_{ab} + \bar h_{\R2\langle ab\rangle} ,\\ 
\hat K^{(2,0)} &= -\frac{31}{6} m^2 a{}_{a} a{}^{a} - \frac{1}{6} \bar h^{a}_{\R2}{}_{a} + \frac{1}{2} \bar h^{tt}_{\R2},
\end{align}
\allowdisplaybreaks
\begin{align}
\hat A^{(2,1)}_a &= \frac{1}{2} \bar h^{b}_{\R2}{}_{b,a} + \frac{1}{2} \bar h^{tt}_{\R2}{}_{,a} 
					- \frac{317}{45} m^2 \mathcal{E}_{ab} a{}^{b} \nonumber\\
					&\quad - \frac{601}{90} m^2 a{}_{a} a{}_{b} a{}^{b} + a{}_{a} \bar h^{b}_{\R2}{}_{b} + 2 a{}_{a} \bar h^{tt}_{\R2},\\
\hat B^{(2,1)} &= -\frac{1}{3} \bar h^{ta}_{\R2}{}_{,a} - \frac{2}{3} a{}^{a} \bar h^{t}_{\R2}{}_{a} + \frac{2}{9} m^2 a{}^{a} a{}_{a}{}_{,t},\\
\hat C^{(2,1)}_{ab} &= -\bar h^{t}_{\R2}{}_{\langle a,b\rangle} + \frac{1}{10} m^2 \dot{\mathcal{E}}{}_{ab} 
					+ \frac{68}{45} m^2 \mathcal{B}_{(a}{}^{d} \epsilon_{b)cd} a{}^{c} \nonumber\\
					&\quad- 2a_{\langle a} \bar h^{t}_{\R2}{}_{b\rangle} + \frac{1}{15} m^2 a_{\langle a} a_{b\rangle}{}_{,t},\\
\hat D^{(2,1)}_a &= -\frac{1}{2} \bar h^{tb,c}_{\R2} \epsilon{}_{abc} + \frac{47}{15} m^2 \mathcal{B}{}_{ab} a{}^{b} 
					+ \epsilon{}_{abc} a{}^{b} \bar h^{t}_{\R2}{}^{c} \nonumber\\
					&\quad + \frac{1}{6} m^2 \epsilon{}_{ab}{}^{c} a{}^{b} a{}_{c}{}_{,t}, \\
\hat F^{(2,1)}_a &= \frac{3}{5} \bar h^{b}_{\R2}{}_{a,b} - \frac{1}{5}\bar h^{b}_{\R2}{}_{b,a} - \frac{67}{90} m^2 \mathcal{E}_{ab} a{}^{b},\\
\hat H^{(2,1)}_{abc} &= \bar h_{\R2\langle ab,c\rangle}- \frac{1}{6} m^2 \mathcal{E}{}_{abc}+ \frac{7}{9}m^2a_{\langle a}\mathcal{E}_{bc\rangle},\\
\hat I^{(2,1)}_{ab} &= -\frac{4}{9} m^2 \dot{\mathcal{B}}{}_{ab} + \frac{2}{3} \bar h_{\R2(a}{}^{c,d} \epsilon_{b)cd} \nonumber\\
					&\quad - \frac{319}{135} m^2 \mathcal{E}_{(a}{}^{d} \epsilon{}_{b)cd} a{}^{c},\\
\hat K^{(2,1)}_{a} &= -\frac{1}{6} \bar h^{b}_{\R2}{}_{b,a} + \frac{1}{2} \bar h^{tt}_{\R2}{}_{,a} 
					- \frac{437}{135} m^2 \mathcal{E}_{ab} a{}^{b} \nonumber\\
					&\quad + \frac{89}{18} m^2 a{}_{a} a{}_{b} a{}^{b} + a{}_{a} \bar h^{tt}_{\R2}.
\end{align}
All quantities on the right-hand side are evaluated at $r=0$.

\vspace{10pt}

We note that the only terms appearing in the second-order equation of motion are $\hat C^{(2,0)}_a$ and $\hat A^{(2,1)}_a$. Because $a^\mu\sim\e$, we have $\hat A^{(2,1)}_a=h^{\R 2}_{tt,i}|_\gamma=\frac{1}{2} \bar h^{b}_{\R2}{}_{b,a}|_\gamma + \frac{1}{2} \bar h^{tt}_{\R2}{}_{,a}|_\gamma+O(\e)$ and $\hat C^{(2,0)}_a = h^{\R 2}_{ta}|_\gamma = - \bar h^{t}_{\R2a}|_\gamma$, showing that for these particular pieces of the two fields, trace-reversing the regular field of definition A yields definition B's regular piece of the trace-reversed field. Therefore in the equation of motion~\eqref{motion_SC} one can make the substitution $h^{\R}_{\mu\nu}=\bar h_{\R \mu\nu}-\frac{1}{2}g_{\mu\nu}g^{\alpha\beta}\bar h_{\R\alpha\beta}$, as one would hope.

\bibliography{second-order}

\end{document}